%% file: main_v3.tex
\documentclass[12pt,twoside,fleqn]{article}
\usepackage{color,colortbl,graphicx,setspace,multirow,enumitem}
\usepackage[utf8]{inputenc}

\usepackage{geometry}
\usepackage[hang,flushmargin]{footmisc}
\usepackage{natbib} 
\usepackage{bibunits} 
\defaultbibliographystyle{custom}
\defaultbibliography{lit}
\usepackage{amsfonts}
\usepackage{amsmath}
\usepackage{amssymb}
\usepackage{placeins}

\usepackage{booktabs,threeparttable} 

\usepackage[dvipsnames]{xcolor}
\definecolor{nblue}{HTML}{000660}
\usepackage[colorlinks=true,urlcolor=nblue,linkcolor=nblue,citecolor=nblue]{hyperref}
\newcommand{\inlinehead}[1]{\noindent{\sffamily\textbf{#1}}. }
\newcommand{\diag}{\text{diag}}
\newcommand{\mtt}[1]{\text{\texttt{#1}}}

\usepackage{titlesec} 
\titleformat{\section}[block]{\bfseries\sffamily\Large}{\thesection. }{0em}{} 
\titleformat{\subsection}[block]{\bfseries\sffamily\large}{\thesubsection. }{0em}{} 
\titleformat{\subsubsection}[block]{\large}{}{0em}{\itshape} 



\usepackage{bm}
\usepackage{caption}
\usepackage{subcaption}
\newcommand\cgiven[1][]{\:#1\vert\:}

\makeatletter\let\p@subfigure\thefigure\makeatother

\usepackage{fancyhdr} 
\makeatletter
\newcommand{\appendixtableofcontents}{\begingroup\par\@starttoc{atoc}\endgroup}
\newcommand{\appsection}[1]{\section{#1}\addcontentsline{atoc}{section}{\protect\numberline{\thesection}#1}}
\newcommand{\appsubsection}[1]{\subsection{#1}\addcontentsline{atoc}{subsection}{\protect\numberline{\thesubsection}#1}}
\makeatother

\def\titletext{Are there asymmetries in euro area monetary policy transmission?}

\title{\sffamily\Huge{\textbf{\titletext}}}
\author{}
\date{}

\begin{document}
\maketitle\vspace{-2.5em}
\normalsize
\begin{center}
\begin{minipage}{.49\textwidth}
  \large\centering Michael \MakeUppercase{Pfarrhofer}\\[0.25em]
  \small WU Vienna
\end{minipage}
\begin{minipage}{.49\textwidth}
  \large\centering Anna \MakeUppercase{Stelzer}\\[0.25em]
  \small Oesterreichische Nationalbank
\end{minipage}
\end{center}

\vspace*{1em}
\doublespacing
\begin{center}
\begin{minipage}{0.85\textwidth}
\noindent We answer the question posed in the title with a nonlinear mixed-frequency vector autoregression, estimated with Bayesian additive regression trees. The model combines monthly macro-financial variables with quarterly bank lending survey data, and identifies the dynamic responses from high-frequency policy surprises. Sign asymmetry dominates; a tightening produces monetary policy transmission mostly in line with the theoretical predictions, whereas easing of any size produces mostly insignificant responses. Peak effects vary with initial conditions, while several common, predefined regime splits do not yield considerable differences in the propagation of the shocks.

\end{minipage}
\end{center}

\singlespacing\vfill\noindent{\footnotesize\textit{Contact}: Anna Stelzer (\href{mailto:anna.stelzer@oenb.at}{anna.stelzer@oenb.at}), Monetary Policy Section, Oesterreichische Nationalbank. \textit{Address}: Otto-Wagner-Platz 3, 1090 Vienna. We thank participants of the 7th ChaMP WS1 workshop, VGSE Research Seminar in Macroeconomics and the NOeG Annual Conference for useful comments and suggestions. We acknowledge the use of Claude Code and Codex for drafting, editing and brainstorming. All AI-assisted suggestions were carefully reviewed, and we take full responsibility for the final content and accuracy of this paper. The views expressed in this paper are those of the authors and do not necessarily reflect the views of the Oesterreichische Nationalbank (OeNB) or the Eurosystem.}

\thispagestyle{empty}
\newpage\doublespacing\normalsize\renewcommand{\thepage}{\arabic{page}}
\renewcommand{\footnotelayout}{\setstretch{1.5}}
\begin{bibunit}

\section{Introduction}\label{sec:introduction}
Understanding how monetary policy transmits to the real economy is central to effective and efficient policy making. Beyond the question of the existence of effects, our paper asks whether and how they depend on the sign of the policy shock, its magnitude, and the distinct economic periods or regimes in which they occur (e.g., the effective lower bound, or recessions). In the euro area, bank lending dominates corporate and household finance, and bank-based transmission channels are particularly important \citep[see, for instance,][]{lane2023}. These aspects are reflected in our modeling approach, which uses a flexible nonparametric mixed-frequency framework, combining lower-frequency survey information on bank lending conditions with standard higher-frequency macroeconomic and financial data.

Bank-based transmission is commonly organized around what \citet{bernanke1995inside} label the credit channel: monetary policy affects not only the general level of interest rates, but also the external finance premium, because credit market frictions can worsen during periods of tighter monetary policy. The balance sheet channel operates through the financial position of borrowers; tighter policy lowers asset values and net worth, making it harder to obtain external financing. The bank lending channel operates through the supply side, as tighter policy reduces banks' capacity to lend.

Empirical studies of the credit channel often focus on heterogeneous transmission due to differences in bank characteristics or across countries \citep{peek1995bank, favero1999transmission, de2013bank, altavilla2020mending}. Our paper adds to another strand of the literature that studies different nonlinearities in monetary policy, among them asymmetries across shock signs and sizes as well as variation over time.\footnote{The cross-sectional heterogeneity of monetary policy transmission across euro area member states \citep[see, e.g.,][for a recent example]{barigozzi2026large} is also referred to as asymmetry. We leave this additional dimension to future research.} \citet{ciccarelli2013heterogeneous} find time variation in bank-based monetary policy transmission in the euro area (EA), and \citet{dahlhaus2017conventional} finds stronger and more persistent effects of expansionary conventional policy in the United States during times of high financial stress. These differences seem to originate from nonlinearities in the balance sheet and bank lending channels. More generally, exploring nonlinearities in monetary policy is not a new endeavor. Studies find time variation in policy, its transmission and the variances of the shocks \citep{primiceri2005time,cogley2005drifts,sims2006were}, as well as dependence on the state of the economy and asymmetries in the sign and size of the shock \citep{weise1999asymmetric,tenreyro2016pushing,barnichon2018functional,alpanda2021state,hauzenberger2021effectiveness,ascari2022non,clark2025nonparametric}. Contractionary policy is typically found to be more potent than expansionary policy, which is sometimes described by the ``pushing on a string'' metaphor \citep[e.g.,][]{angrist2018semiparametric}.

Much of the existing evidence faces two limitations. First, many studies either rely on conditionally linear models that impose symmetric transmission, or specify a single type of nonlinearity (a particular regime split, threshold, or transition function) by design of the econometric model, so that the form of any asymmetry is chosen ex ante rather than inferred. Second, high-frequency monetary policy surprises need to be aggregated when used with lower-frequency data, averaging out within-quarter variation in the shock series and possibly inducing temporal aggregation biases. Given the relative importance of bank-based financing in the euro area, and the fact that survey information on bank lending is available only at a quarterly frequency, we address both issues with a nonparametric model that does not impose a particular sign or size response function, in a mixed-frequency setup that uses the shock at the monthly frequency.\footnote{Mixed-frequency models are well suited to alleviate temporal aggregation biases \citep{foroni2016mixed}. \citet{jacobson2023temporal} relate these biases to common wrong-signed impact estimates of inflation in a days-to-months frequency mismatch; our approach does not address that finer mismatch, but it avoids adding a month-to-quarter aggregation of the shock on top of it.}

Alongside standard macroeconomic and financial variables at both quarterly and monthly frequency, we use information on bank-based transmission channels in the EA, as captured by data of the Bank Lending Survey (BLS). This is a quarterly survey by the Eurosystem that reports bank lending conditions since $2003$. Inspired by \citet{ciccarelli2013heterogeneous}, we summarize the information contained in the survey into common factors to characterize the broad credit channel, the bank lending channel and the borrower balance sheet channel. These factors are survey measures of the conditions through which the respective channels operate.

Methodologically, we propose a multivariate nonparametric mixed-frequency model in which the conditional mean functions are estimated with Bayesian Additive Regression Trees (BART), building on \citet{huber2023nowcasting} and \citet{marcellino2025nonparametric}. The key addition relative to these papers is that an observed high-frequency monetary policy shock enters the conditional mean functions contemporaneously.\footnote{BART is also used with local projections to estimate GIRFs \citep{mumtaz2022impulse}.} The shock is constructed from surprises around frequent monetary policy events --- Governing Council decisions as well as speeches by the European Central Bank (ECB) President and Executive Board members --- taken from the extended event-study database of \citet{altavilla2025monetary}, and purged of central bank information effects with the sign restrictions of \citet{jarocinski2020deconstructing}. Under the assumption that the resulting shock series is exogenous, we compute generalized impulse responses (GIRFs) to the shock and capture nonlinearities arising on impact and as the shock propagates through the economy.

Our results yield three main findings. First, the responses differ by the sign of the shock, which corroborates a literature in which contractionary policy is typically found to be more impactful than expansionary policy. Contractionary shocks yield significant effects already at $1$ standard deviation (sd) for stock prices, corporate spreads and unemployment; aggregate output and the price level decline, with marginally significant responses. Expansionary shocks produce mostly insignificant responses. A kinked form is the most frequent descriptive approximation of the estimated impact functions for every variable, flat for easing and sloped for tightening for the credit factors, the corporate spread and stock prices. These curves also illustrate what a linear estimator applied to the same shock would report, namely a weighted average of the flat easing and the active tightening side with weights determined by the distribution of the shock \citep{kolesar2024dynamic}.

Second, the survey-based credit standards and borrower balance sheet factors tighten persistently after contractionary shocks. These factors capture a dimension of transmission that the estimated muted impact response of the short-term rate to tightening would miss. Third, comparing credible bands across recessions and expansions, at and away from the effective lower bound (ELB), and across positive and nonpositive credit-to-GDP gaps (as a measure for financial imbalances), we find no separation, except for the response of a short-term interest rate at the ELB after larger tightenings. At the same time, median responses at propagation horizons vary with the initial macro-financial conditions from which a shock occurs, in some periods by a factor of two, a time variation not captured by the predefined regimes.

The remainder of the paper is structured as follows. Section \ref{sec:data} describes the data, the monetary policy shock and the empirical design, Section \ref{sec:econometrics} our econometric framework, and Section \ref{sec:empirics} our empirical findings. We conclude in Section \ref{sec:conclusions}. The Online Appendix contains additional details and results.

\section{Data and empirical design}\label{sec:data}
Our mixed-frequency econometric framework is designed for the euro area, with a specific data environment in mind. Before turning to the modeling approach, we thus present the dataset and the empirical design, including how we construct the monetary policy shock and the regime classifications.

\subsection{Variables and credit-condition factors}\label{sec:variables}
The monthly block contains typical variables for a monetary analysis, including industrial production, the unemployment rate, the harmonized index of consumer prices (HICP), the 3-month and 1-year Euribor, a 10-year sovereign yield, the Euro Stoxx 50, a euro high-yield corporate bond option-adjusted spread (OAS) as a measure of financial conditions and the nominal effective exchange rate. The quarterly block contains real GDP and four factors extracted from the BLS. Appendix \ref{app:empirical} lists sources and transformations. When reporting responses, we cumulate differenced series and divide annualized series by their underlying annualization factor. Responses are reported in percent for real GDP, industrial production, stock prices, the exchange rate and the HICP; basis points for the interest rates and the spread; percentage points for the unemployment rate; and standard deviations of the respective factor for the BLS factors, obtained by dividing their responses by the sample standard deviation of the factor. The HICP response is the cumulative response of the price level. The multivariate time series model runs at the monthly frequency and the data span January $1999$ to December $2024$; with $p = 12$ lags, the estimation sample covers $T = 300$ months.

\begin{figure}[ht]
    \centering
    \includegraphics[width=\linewidth]{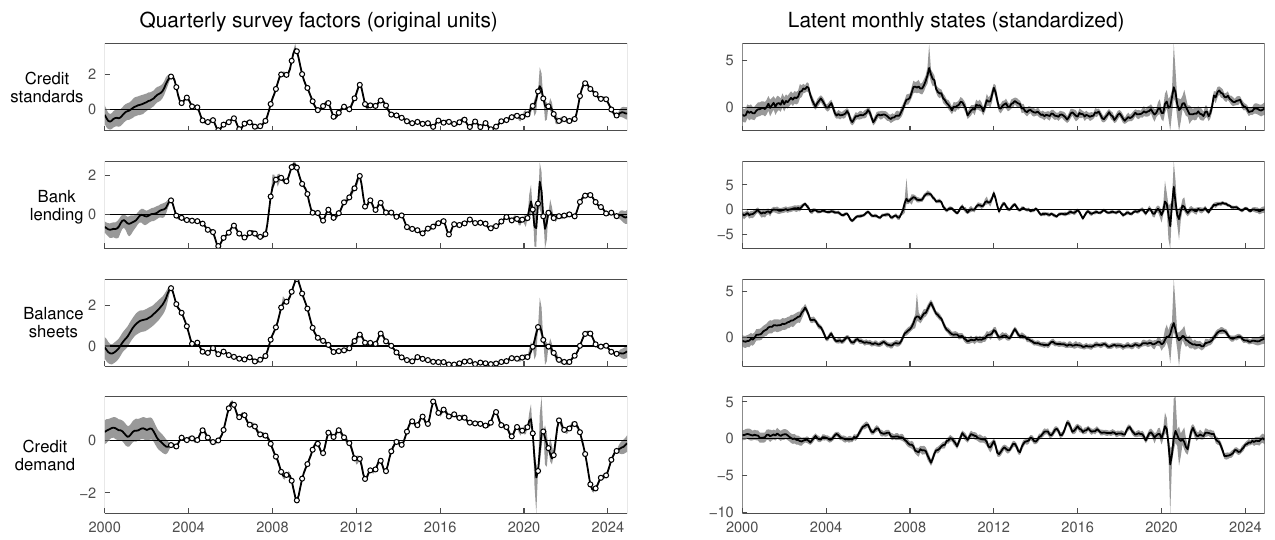}
    \caption{Survey-based credit-condition factors, with model-implied monthly states.}
    \label{fig:factors}\vspace*{-0.25cm}
    \caption*{\footnotesize \textit{Notes}: Left: quarterly factors in original units; circles mark the observed survey values ($2003$Q$1$ to $2024$Q$2$), the line and band the model-implied quarterly series (median and $68$ percent band, baseline specification), which is inferred outside survey coverage. Right: latent monthly states (standardized units).}
\end{figure}

The BLS is a quarterly survey of senior loan officers at a representative sample of euro area banks, available from $2003$Q$1$; our survey sample ends in $2024$Q$2$, so the factors cover $86$ quarters while real GDP is observed for $104$ quarters. Each survey item reports a net percentage of banks that tightened rather than eased credit standards (or that observed an increase rather than a decrease in demand) over the past three months, so the underlying series describe \textit{changes} in credit conditions. Following \citet{ciccarelli2013heterogeneous}, we extract the first principal component of four blocks of items. The \textit{broad credit} factor (labeled \texttt{BC}) summarizes changes in credit standards for loans to enterprises, for house purchase and for consumer credit; the \textit{demand} factor (\texttt{D}) summarizes changes in loan demand for the same three loan categories; the \textit{bank lending} factor (\texttt{BL}) summarizes the bank-side factors that loan officers cite as contributing to changes in standards (capital costs, market funding, liquidity, competition); and the \textit{borrower balance sheet} factor (\texttt{BBS}) summarizes borrower-side factors (general and industry-specific outlook, collateral risk, creditworthiness). Appendix \ref{app:empirical} lists the items in each block. The factors are normalized so that an increase indicates tighter credit standards or, for the demand factor, higher loan demand. Because the survey starts in $2003$ and ends in mid-$2024$, $42$ of the $300$ months of the effective sample lie outside the observed survey coverage; for these months, the credit conditions are inferred by the model, and full-sample averages of responses include these model-inferred histories.

Figure \ref{fig:factors} shows the four factors (left panels), together with the latent monthly states that our model infers for them (right panels). The properties of these series partially motivate our modeling choices. First, credit conditions move episodically. The supply-side factors are close to zero for long stretches and spike during the global financial crisis, the sovereign debt crisis, the pandemic and the tightening cycle of $2022$--$2023$, whereas easing phases are more gradual. The raw series thus already display an asymmetry between abrupt tightening and slow easing. Second, the supply-side factors co-move strongly and loan demand mirrors them. But they also show idiosyncrasies, e.g., the recent tightening cycle is visible in credit standards, bank lending and demand but slightly less so in borrower balance sheets. Modeling the factors jointly preserves such more nuanced information. Third, the factors are quarterly and start only in $2003$, whereas the shocks and the financial variables are monthly, which motivates the mixed-frequency design that infers the monthly path of credit conditions rather than aggregating the shock to the quarterly frequency.

\subsection{The monetary policy shock and regimes}
The monetary policy shock, encoded as $\varepsilon_t$, is constructed from high-frequency overnight index swap (OIS) surprises around frequent monetary policy events. We use the Euro Area Extended Monetary Policy Event-Study Database \citep[EA-EMPD,][]{altavilla2025monetary}, which extends the Governing Council event windows of \citet{altavilla2019measuring} to speeches by the ECB President and by Executive Board members. We include the monetary event window of each Governing Council meeting (press release and press conference combined) and the speech windows. We extract the first principal component of the OIS surprises at maturities between 1 month and 1 year across all events, and we treat this principal component, $s_{\tau}^{\text{(OIS)}}$, as our policy rate surprise for event $\tau$.

\begin{figure}[ht]
    \centering
    \includegraphics[width=\linewidth]{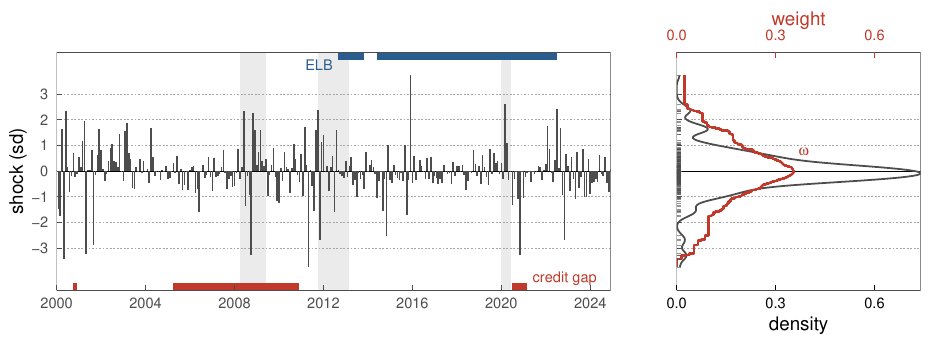}
    \caption{Monetary policy shock series and regime classifications.}
    \label{fig:shockts}\vspace*{-0.25cm}
    \caption*{\footnotesize \textit{Notes}: Standardized monthly shock, January $2000$ to December $2024$. Gray shading: recessions; blue bar (top): ELB; red bar (bottom): positive credit-to-GDP gap. Right: density (gray) and realized values, and the weight function $\omega$ (red), which describes how linear estimators would average marginal effects.}
\end{figure}

Next, we use the corresponding surprise in equity prices $s_{\tau}^{\text{(ES50)}}$ to separate monetary policy shocks from central bank information effects with the approach of \citet{jarocinski2020deconstructing}. That is, we assume $(s_{\tau}^{\text{(OIS)}}, s_{\tau}^{\text{(ES50)}})' = \bm{B}_0^{-1} (\varepsilon_{\tau}, \tilde{\varepsilon}_{\tau})'$, and impose the sign restrictions on $\bm{B}_0^{-1}$ such that a monetary policy shock moves the OIS and equity surprises in opposite directions, while an information shock moves them in the same direction. We then recover the monetary policy shock by pre-multiplying the vector of surprises with $\bm{B}_0$. The resulting series is aggregated by summing over events within a month, purged of first-order autocorrelation, standardized to unit variance, and displayed in Figure \ref{fig:shockts}. A positive value is a tightening shock. Its correlation with the monthly sum of the raw OIS factor is $0.68$, and a $1$ sd shock corresponds to a monthly surprise of about $3$ basis points in the 3-month and 1-year OIS rates. Appendix \ref{app:empirical} provides additional details and results for each intermediate step.\footnote{Note that we condition on the median estimate of the shock, so the uncertainty from this step is not propagated.}

The sample contains $160$ easing and $140$ tightening months of similar average absolute size ($0.60$ and $0.64$ sds). The tails are sparse, with nine months below $-2$ sds and seven months above $+2$, and several of the largest easing shocks fall in crisis months. We later compute responses to shocks of $d \in \{\pm1, \pm2, \pm3\}$ sds; $55$ months have $|\varepsilon_t| \geq 1$, against $16$ with $|\varepsilon_t| \geq 2$ and six with $|\varepsilon_t| \geq 3$ (five of them easings), so the larger scenarios rest on sparse tails.

A useful diagnostic in a nonlinear context is the weight function described in \citet{kolesar2024dynamic}, who show that a linear estimate on an observed shock in a nonlinear data generating process identifies a weighted average of the marginal effects across baseline values $x$ of the shock. The weight function $\omega(x) = \mathrm{Cov}(\mathbb{I}(\varepsilon_t \geq x), \varepsilon_t)/\mathrm{Var}(\varepsilon_t)$ is nonnegative, integrates to one, is hump-shaped around the mean of the shock and depends only on its marginal distribution. Estimated from our shock series and shown in the right panel of Figure \ref{fig:shockts} next to the density, the weight function places $47$ percent of its mass on tightenings, $56$ percent on shocks within $1$ sd and $84$ percent within two, and only $5$ and $12$ percent beyond $+2$ and $-2$ sds. A linear summary of the effects of this shock series would thus be dominated by small shocks of either sign; by contrast, our approach explicitly estimates nonlinearities at distinct shock magnitudes.

\inlinehead{Regime classifications} We consider three predefined binary regime classifications, shown in Figure \ref{fig:shockts}: (1) recessions as dated by the Euro Area Business Cycle Dating Committee ($39$ months) versus expansions; (2) the ELB, defined as months in which the 3-month Euribor was below $25$ basis points ($113$ months, from September $2012$ to July $2022$ with a short interruption), similar to \citet{carriero2025forecasting}; and (3) months with a positive credit-to-GDP gap ($81$ months, between October $2000$ and March $2021$) versus months with a nonpositive gap. The credit-to-GDP gap is the ECB's quarterly series carried forward to months, available from October $2000$ (earlier months count as nonpositive); a positive gap is an early-warning indicator of credit expansion relative to trend. Both signs of the shock occur in every regime. Recessions contain $16$ easing and $23$ tightening months, ELB months $71$ and $42$, and positive-gap months $47$ and $34$. Easing shocks are somewhat overrepresented at the ELB, but otherwise these shares are roughly balanced. 

The shocks in the tails are split across regimes more sparsely. Of the $16$ months with $|\varepsilon_t| \geq 2$, recessions contain six (two easings, four tightenings), ELB months five (two and three) and positive-gap months four (two and two), and the only tightening above $3$ sds falls in an ELB month outside recessions and positive-gap periods. Shock impacts beyond about $2$ sds are thus estimated from very few observations or extrapolated from smaller shocks, and should be interpreted with this in mind.

\section{Econometric framework}\label{sec:econometrics}
\subsection{A nonlinear mixed-frequency model}\label{sec:econometrics_mfvar}
Let $\bm{y}_t = (\bm{y}_{\mtt{l},t}',\bm{y}_{\mtt{h},t}')'$ be the $n\times1$ vector of monthly variables, where $\bm{y}_{\mtt{h},t}$ collects the $n_{\mtt{h}}$ variables observed each month and $\bm{y}_{\mtt{l},t}$ the latent monthly counterparts of $\bm{y}_{\mtt{l},t}^{\ast}$, the $n_{\mtt{l}}$ variables observed only quarterly, so that $n = n_{\mtt{h}} + n_{\mtt{l}}$. Let $\bm{z}_t = (\bm{y}_{t-1}',\hdots,\bm{y}_{t-p}')'$ contain $p$ lags, and let $\varepsilon_t$ denote the observed monetary policy shock. Finally, $s_t \in \{1,\hdots,S\}$ indexes the $S$ discrete regimes that are known a priori. With $\mathbb{I}(\bullet)$ denoting the indicator function, the most general form of our state equation is:
\begin{equation}
    \bm{y}_t = \sum_{s=1}^{S}\left[\bm{H}_s(\bm{z}_t) + \bm{G}_s(\varepsilon_t) + \bm{O}_t\bm{u}_{st}\right] \cdot \mathbb{I}(s_t = s), \quad \bm{u}_{st} \sim \mathcal{N}(\bm{0}_n, \bm{\Sigma}_{s}),\label{eq:MFBART}
\end{equation}
where $\bm{H}_s(\bm{z}_t) = (h_{s1}(\bm{z}_t),\hdots,h_{sn}(\bm{z}_t))'$ and $\bm{G}_s(\varepsilon_t) = (g_{s1}(\varepsilon_t),\hdots,g_{sn}(\varepsilon_t))'$ are regime- and equation-specific unknown functions, $\bm{\Sigma}_s$ is a regime-specific covariance matrix, and $\bm{O}_t = \diag(o_{1t},\hdots,o_{nt})$ collects variable-specific outlier scalings in the spirit of \citet{carriero2021addressing}. We estimate all $h_{si}$ and $g_{si}$ with BART, so both the propagation of the system and the contemporaneous effect of the shock are inferred nonparametrically. The model is estimated on data standardized to zero mean and unit variance; all reported results are scaled to original units ex post.

Because the model is nonlinear, the response to a shock generally depends on the macro-financial conditions from which it occurs, that is, on the initial conditions stored in $\bm{z}_t$; see also the definition of the GIRF in Equation (\ref{eq:definitionIRF}). Within a regime $s$, the contemporaneous effect of a shock depends only on the shock itself through $\bm{G}_s$, so the impact response is the same regardless of these conditions; its shape as a function of the shock, however, is unrestricted, so sign asymmetries and size nonlinearities on impact are estimated freely. From the first month after impact onward, distinct initial conditions can change the response, because the propagation through $\bm{H}_s$ is nonlinear; this may introduce time variation in the dynamic responses even though $\bm{H}_s$ itself is constant within a regime. In multi-regime models, both $\bm{G}_s$ and $\bm{H}_s$ may differ across regimes, so the impact and the dynamic responses can both be regime-dependent and thus vary over time.

We treat the monetary policy shock $\varepsilon_t$ as observed, as is common in the literature in the spirit of \citet{romer2004new,gurkaynak2005actions}. We assume that, conditional on the regime $s_t$, it is jointly independent of the history $\bm{z}_t$ and any other structural shocks affecting the economy in period $t$ and thereafter.\footnote{The independence assumption concerns the underlying structural shocks. Outcomes depend on the monetary policy shock through the model's response functions. Removing autocorrelation and separating the monetary policy shock from information effects address related threats to identification.} Under this assumption, the response comparisons below have a causal interpretation.

The quarterly observations $y_{\mtt{l},it}^{\ast}$, $i = 1,\hdots,n_{\mtt{l}}$, are linked to the latent monthly series through approximate measurement equations:
\begin{equation}
y_{\mtt{l},it}^{\ast} = \frac{1}{3}\left(\frac{1}{3} y_{\mtt{l},it} + \frac{2}{3} y_{\mtt{l},it-1} + y_{\mtt{l},it-2} + \frac{2}{3} y_{\mtt{l},it-3} + \frac{1}{3} y_{\mtt{l},it-4}\right) + \eta_{it}, \label{eq:MFBART_ME}
\end{equation}
in every third month, in the spirit of mixed-frequency VARs. The triangular weights are the intertemporal restriction for a quarterly period-over-period change built from monthly changes; they rest on the log-linear approximation of \citet{mariano2003new}, under which quarterly log growth is approximately a weighted sum of monthly log growth rates, and are therefore an approximation rather than an identity. Observed real GDP enters as annualized quarterly log growth, and because the weights sum to one, the latent monthly series is on the same annualized scale. When reporting responses, the latent monthly response is cumulated and divided by four, the annualization factor of the observed quarterly series. For the BLS factors, the same weights are used. Following \citet{chan2023high}, we introduce measurement errors $\eta_{it} \sim \mathcal{N}(0, \omega_{i}^2)$ with a small variance $\omega_{i}^2 = 10^{-8}$.

\subsection{Nonparametric conditional means}\label{sec:bart}
BART \citep{chipman2010bart} approximates the unknown conditional mean and shock impact functions by a sum of $B$ regression trees, $h_{si}(\bm{z}_t)\approx \sum_{b=1}^{B}\ell_{si,b}(\bm{z}_t|\mathcal{T}_{si,b}, \bm{\mu}_{si,b})$, where $\mathcal{T}_{si,b}$ is the structure of tree $b$ and $\bm{\mu}_{si,b}$ its terminal node parameters; the shock functions $g_{si}(\varepsilon_t)$ are approximated in the same way with the shock as the only input. We use $B = 250$ trees per function and the default prior of \citet{chipman2010bart} on the tree structure, with terminal node priors calibrated to the observed range of each variable within each regime. When $S = 1$ the model is close to the multivariate BART models of \citet{huber2020inference,huber2023nowcasting,marcellino2025nonparametric}. Because $\bm{H}_s$ is a nonlinear function of the latent monthly states, we follow \citet{marcellino2025nonparametric} and sample the latent states from a shrinkage-based linear approximation of the fitted trees within a Gibbs sampler. The prior setup follows default choices in the BART literature, see also Appendix \ref{app:technical}.

\subsection{Impact responses and dynamic causal effects}\label{sec:GIRFdescription}
We use two related objects to describe the effects of the monetary policy shocks. The first is the estimated shock function itself. For a grid of shock values $\varepsilon_l$, the impact curve $g_{si}(\varepsilon_l) - g_{si}(0)$ shows the contemporaneous response of variable $i$ to a shock of size $\varepsilon_l$ relative to no shock. This curve, a partial dependence plot in machine-learning jargon, has a zero baseline and does not depend on initial conditions by construction. The second object is the dynamic response. Following \citet{koop1996impulse}, we define the generalized impulse response of variable $i$ at horizon $h$ as the difference between two conditional expectations,
\begin{align}
    \text{I$\tilde{\text{R}}$F}(d,\bm{\Omega}_t,s)_{i,h} = \mathbb{E}(y_{i,t+h} \cgiven \varepsilon_{t} = d_0 + d,\bm{z}_t, s_t = s) -\mathbb{E}(y_{i,t+h} \cgiven \varepsilon_{t} = d_0,\bm{z}_t, s_t = s),\label{eq:definitionIRF}
\end{align}
where the shock in period $t$ is set to a baseline value $d_0$ in the second term and to $d_0 + d$ in the first, $d$ encodes the sign and size of the experiment, and $\bm{\Omega}_t = (d_0,\bm{z}_t')'$ collects the conditioning arguments, which we refer to below as the full set of initial conditions.\footnote{For a taxonomy of dynamic causal effects, see \citet{rambachan2021common,gonccalves2022state}.} The baseline used below is $d_0 = 0$. The expectations in (\ref{eq:definitionIRF}) integrate over the shocks and reduced-form innovations in periods $t,\hdots,t+h$. We evaluate them by simulation. The initial conditions $\bm{z}_t$ are the $p$ months preceding period $t$ in the (latent-augmented) sample, and period $t$ itself is simulated with its shock replaced by $d_0$ or $d_0 + d$. Future shocks are drawn with replacement from the realized shock series, and innovations from their Gaussian distribution; the same random draws are used along the shocked and the baseline path, which reduces Monte Carlo simulation noise in the difference. The regime of the origin period is held fixed along the simulated path, so the responses describe transmission within a regime rather than across regime transitions. Responses reported for the full sample or for a regime average the period-specific responses over the corresponding dates within each posterior draw, so that the credible bands reflect posterior uncertainty about the averaged response. Appendix \ref{app:technical} provides additional details.

To make experiments of different signs and sizes comparable, we mostly report the normalized response $\text{IRF}(d,\bm{\Omega}_t,s)_{i,h} = \text{I$\tilde{\text{R}}$F}(d,\bm{\Omega}_t,s)_{i,h} / d$, expressed per $1$ sd of tightening. Under proportional transmission the normalized responses to all $d$ coincide; differences across $|d|$ within a direction indicate size nonlinearity, and differences between $d$ and $-d$ indicate sign asymmetry. Most of our aggregate results show time averages, so we report $\text{IRF}(d,s)_{i,h} = \sum_{t=1}^{T}\text{IRF}(d,\bm{\Omega}_t,s)_{i,h} / T$, with $T$ the number of months in the sample or in the regime. The later sections show $t$-by-$t$ estimates where variation can emerge across initial conditions and across regimes.

\section{Asymmetric effects of monetary policy} \label{sec:empirics}
We start with the results for the model defined in (\ref{eq:MFBART}) and (\ref{eq:MFBART_ME}) for a single regime $S = 1$. We characterize the shape of the impact responses, then summarize the dynamic responses by their peaks across the sign and size of the shocks, turn to the full dynamic responses, and finally explore variation over time and across $S>1$ regimes.

\subsection{Which types of asymmetries exist on impact?}\label{sec:impactasym}
\begin{figure}[ht]
    \centering
    \includegraphics[width=\linewidth]{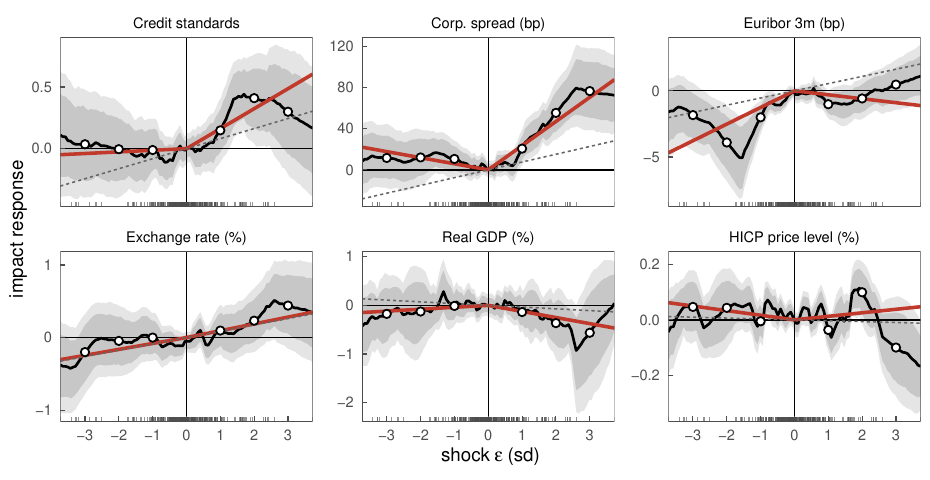}
    \caption{Nonlinear impact effects as a function of the shock $\varepsilon$ for selected variables.}
    \label{fig:shock_vars}\vspace*{-0.25cm}
    \caption*{\footnotesize \textit{Notes}: Basis points (bp), percentage points (pp). Posterior median with $50/68$ percent credible sets; red: kink form $\beta_1\varepsilon + \beta_2|\varepsilon|$; dashed gray: linear form $\beta_1\varepsilon$; both through-origin fits to the median at the $300$ realized shocks, so the linear line is the slope a linear estimator on the shock would report. Circles: median impact at $\varepsilon \in \{\pm1,\pm2,\pm3\}$. Rug: realized shocks.}
\end{figure}

Figure \ref{fig:shock_vars} shows the nonlinear impact effect on six selected variables as a function of the shock over a grid of values $\varepsilon_l$ spanning the observed support, i.e., the impact curve. These partial dependence plots show the shape of the functional relationship between the shock and the respective variable on impact, and the selection spans the shapes found across the system. The circles mark the median impacts at the shock values $\varepsilon \in \{\pm1,\pm2,\pm3\}$, i.e., the impacts from which the dynamic responses of Section \ref{sec:dynamics} originate. The dashed gray line is the linear approximation of each curve, fitted at the values of the realized shocks. It represents a weighted average of the marginal effects along the curve, with weights that depend only on the distribution of the shock (see also Section \ref{sec:data}). The curves for the remaining variables are provided in Appendix \ref{app:addresults}.

  To characterize these shapes quantitatively, we approximate the nonparametric estimate by three parametric references \citep[see also][]{caravello2024disentangling}: linear, $\beta_{i1}\varepsilon_l$; sign nonlinearity (a kink at zero), $\beta_{i1}\varepsilon_l + \beta_{i2}|\varepsilon_l|$; and size nonlinearity (we do not show this approximation in the figures because of its empirically subordinate importance), $\beta_{i1}\varepsilon_l + \beta_{i2}\varepsilon_l|\varepsilon_l|$. We estimate each form by OLS on every posterior draw, evaluated at the observed shock realizations in the estimation sample, and classify each draw by the form with the lowest BIC. Table \ref{tab:gxfits_full} reports $68$ percent credible sets of the slope of the linear form, of the easing and tightening slopes of the sign-shape and of their difference, together with the shares of draws in which each form attains the lowest BIC. These parametric forms are descriptive approximations, and the winner shares represent a rough gauge of best fit.

The sign nonlinearity is the best-fitting approximation for all $14$ variables, by $56$ to $41$ percent for unemployment at the narrowest and by $95$ to $4$ percent for the corporate spread at the widest. The linear form's share never exceeds $16$ percent. The strength and significance of this pattern are heterogeneous across variables. The slope difference excludes zero for the corporate spread, the stock market and the 3-month rate. For the corporate spread, the easing-side slope has a set that includes zero, while the tightening-side slope is positive with a set that excludes it. The stock market shares this pattern with the opposite sign, while the 3-month rate displays a reverse orientation --- an easing-side slope whose set excludes zero and a tightening-side slope whose set includes it, at a small magnitude. We return to this somewhat puzzling result at the end of this subsection. The exchange rate is close to linear in shape, and for output, industrial production, unemployment and the price level both slopes feature wide credible sets. Significant tightening impact estimates emerge for the Euro Stoxx 50, corporate spreads, and the BLS factors except credit demand.

\begin{table}[t]
\caption{Parametric approximations of the estimated impact functions.}\label{tab:gxfits_full}\vspace*{-1em}
\begin{center}
\begin{threeparttable}
\scriptsize\setlength{\tabcolsep}{1.8pt}
\input{plots_paper/tab_gx_fits.tex}
\begin{tablenotes}[para,flushleft]
\footnotesize{\textit{Notes}: OLS fits to each posterior draw of the baseline impact function at the $300$ observed shocks. Units per sd of the shock: basis points (bp), percent, and percentage points (pp). Entries are $68$ percent credible sets across posterior draws; $^{*}$ marks a set excluding zero. Linear: slope of $\beta_1\varepsilon$. Sign: $\beta_1\varepsilon + \beta_2|\varepsilon|$, whose easing slope is $\beta_1 - \beta_2$ and tightening slope $\beta_1 + \beta_2$; size: $\beta_1\varepsilon + \beta_2\varepsilon|\varepsilon|$. BIC winner shares (rounded) give the percentage of draws in which each of the three forms (linear, sign, size) has the lowest BIC.}
\end{tablenotes}
\end{threeparttable}
\end{center}
\end{table}

The linear column of Table \ref{tab:gxfits_full} contains the slopes a linear estimator applied to these shocks would report; they are average marginal effects along the curve under the weight function of \citet{kolesar2024dynamic}, and draw by draw they equal the easing and tightening slopes weighted by each side's share of the sum of squared shocks. These weighted averages exhibit the signs that standard theory suggests for a monetary tightening (and easing proportionally, by construction, in a linear world). Credit standards tighten, spreads widen, stock prices fall, and output, prices and credit demand decline, with sets that exclude zero for credit standards, the spread, stock prices and industrial production. The nonparametric curves show where these averages come from. Some segments of the curves over the support of the shock are puzzling, most visibly in the tails, and these segments are informed by a small number of months. The next paragraph discusses these in more detail.

In the far easing tail, several curves revert toward tightening patterns. Real GDP in Figure \ref{fig:shock_vars} responds negatively to a $3$ sd easing, and unemployment (shown in Appendix \ref{app:addresults}) responds positively, whereas for a $1$ sd easing unemployment has the conventional sign and the output response is zero. The tightening tail occasionally shows similar features, for instance for the credit standards factor above shock size $2$. The associated bands, however, typically cover zero. This can be linked mechanically to how BART fits data. The shock function is a sum of trees with the shock as the only input, so it is a step function whose value on an interval is an average over the months whose shocks fall into that interval. Beyond $2$ sds these are $16$ months, beyond $3$ sds six months, five of them easings and a single tightening. \citet{gouletcoulombe2025blackbox} show that tree-based impulse responses aggregate past outcomes with sparse proximity weights and document for financial shocks that the asymmetric effects found by such estimators trace back to a few identifiable distress episodes. The tails of our impact curves are likewise averages over a handful of crisis months, and the puzzling estimates are averages of what happened in those months. A linear estimator would hardly register this region, since the weight function of Section \ref{sec:data} assigns only $5$ and $12$ percent to it; the nonparametric estimate exposes some of these oddities.

The impact response of interest rates is small, with all interest rate slopes in Table \ref{tab:gxfits_full} within a few basis points per sd, and it is asymmetric in the opposite direction to the financial and credit variables. The easing slope of the 3-month Euribor, about $1.3$ basis points, has a set that excludes zero, whereas the tightening slope is insignificant. In the linear approximation the 3-month Euribor rises by about $0.6$ basis points with a set that excludes zero, but this slope is driven by the easing side, and no rate rises after a tightening with a set that excludes zero. The corporate spread, by contrast, widens after tightenings, with a slope of $15$ to $33$ basis points per sd ($68$ percent set), so tightening shocks raise the cost of risky borrowing even though the short rate does not move. This is in line with the evidence for the US in \citet{gertler2015monetary}, who find that policy shocks produce modest movements in short rates but large movements in credit costs, mainly through term premia and credit spreads. Despite the insignificant response of short-term rates to tightenings, the survey-based credit factors as well as the macroeconomic and financial variables react, both on impact and at propagation horizons. Because the factors we extract from the BLS are based more broadly on credit standards, rather than narrowly on interest rates, they capture a dimension of transmission that the insignificant impact response of the 3-month Euribor to tightenings would miss.\footnote{The lack of a significant positive reaction of short-term rates to contractionary surprise-based shocks in the euro area has been documented also in \citet{jarocinski2020deconstructing}, who, unlike in their model for the US, impose an additional sign restriction on the corresponding monthly interest rate.}

\subsection{An aggregate view: peak responses across sign and size}\label{sec:aggregate}
\begin{figure}[p]
    \centering
    \includegraphics[width=\linewidth]{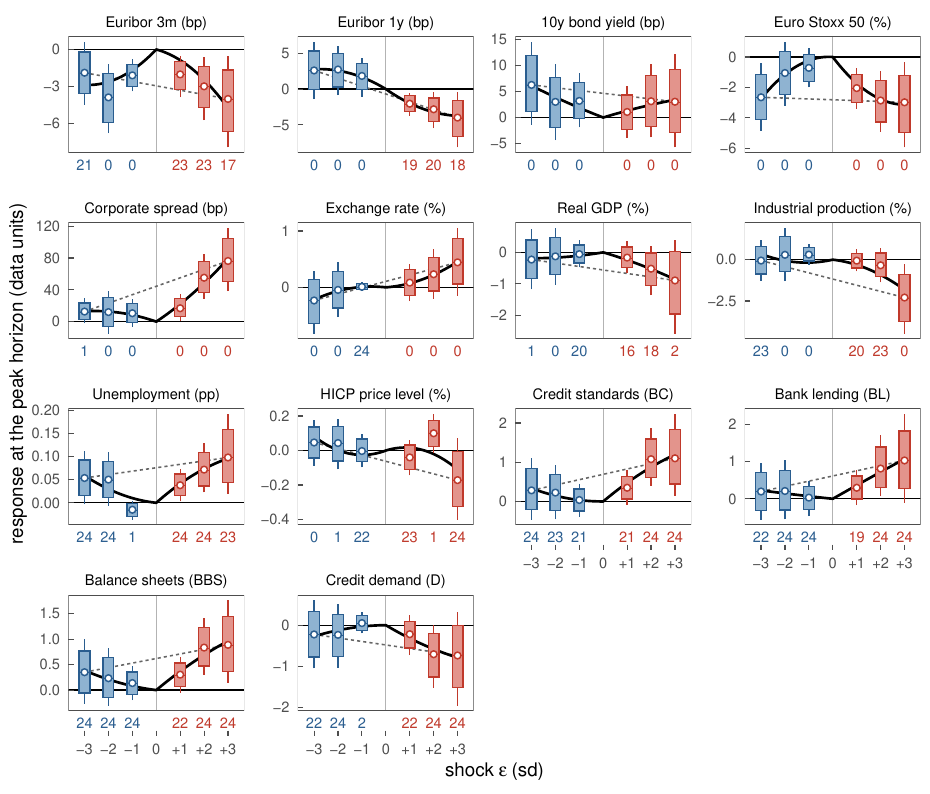}
    \caption{Peak responses as a function of the size and sign of the shock.}\label{fig:peaksize}\vspace*{-0.25cm}
    \caption*{\footnotesize \textit{Notes}: Baseline specification, data units (not divided by the shock size; units as in Figures \ref{fig:main-irfs_credit}--\ref{fig:main-irfs_macro} and the corresponding response charts in Appendix \ref{app:addresults}). For each shock $d \in \{\pm1, \pm2, \pm3\}$ sd, the peak horizon (printed in a separate strip beneath each panel) is where the absolute posterior median is largest; circles, boxes and whiskers are the median, $50$ and $68$ percent credible sets of the response at that horizon. Gray dashed: line through the $\pm3$ sd medians; black: $\beta_1 d + \beta_2 |d| + \beta_3 d|d|$ fitted through the six medians. A kink at zero indicates sign asymmetry, curvature within a direction size nonlinearity.}
\end{figure}

Before turning to the full dynamic responses, Figure \ref{fig:peaksize} provides a summary. The figure shows the response at the peak horizon, together with its $50$ and $68$ percent credible sets. Proportional and non-state-dependent transmission would correspond to identical timings of the peaks and a straight line through the origin.

The sign dimension is again the more clear-cut of the two. For the credit standards, bank lending and borrower balance sheet factors, the corporate spread, stock prices and unemployment, the tightening side moves away from zero with the size of the shock while the easing side stays flat, with credible sets that include zero at every easing size; the sole exception is the stock market after a $3$ sd easing, with a wrong-signed response from a theory perspective. Comparing responses of the same absolute size but opposite sign, the peak responses to tightening of unemployment, stock prices and the spread (at every size), of the credit standards and borrower balance sheet factors (at $2$ and $3$ sds), of the bank lending factor (at $2$ sds) and of industrial production (at $3$ sds) have $68$ percent sets that exclude zero, whereas none of the corresponding easing responses does. We interpret this as evidence of sign asymmetry.

The size dimension is less clear. For the credit standards and borrower balance sheet factors, the peak responses to tightening rise from $0.35$ and $0.30$ standard deviations for a $1$ sd shock to $1.08$ and $0.83$ for two, but only to $1.10$ and $0.89$ for three, so the response per sd is largest at $2$ sds and the $68$ percent bands of the three sizes overlap at every horizon (Figure \ref{fig:main-irfs_credit}). Among the remaining variables, the spread and unemployment scale roughly proportionally with the shock, the stock market flattens out beyond $2$ sds, industrial production responds with a set excluding zero only at the largest tightening, and the price level is not monotone in the shock size. The short rates rise modestly after a tightening; about one and a half years later, they are below the baseline, consistent with policy responding to the weaker outlook.

\subsection{The dynamic picture}\label{sec:dynamics}
\begin{figure}[ht]
    \centering
    \includegraphics[width=\linewidth]{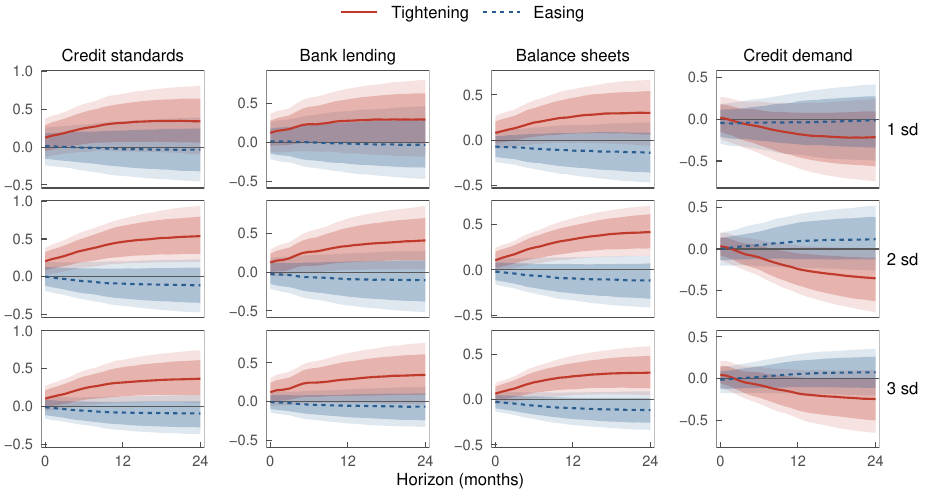}
    \caption{Impulse response functions of the credit variables to monetary policy shocks of different sign and size.}\label{fig:main-irfs_credit}\vspace*{-0.25cm}
    \caption*{\footnotesize \textit{Notes}: Baseline specification, averaged over all initial conditions. Rows: absolute shock sizes of $1$, $2$ and $3$ sd; columns: variables. Each panel pairs tightening (red, solid median) and easing (blue, dashed median), with darker $50$ and lighter $68$ percent credible bands. Standard deviations of the respective factor, normalized per $1$ sd of tightening (Section \ref{sec:GIRFdescription}), so a conventionally signed easing response has the same sign as the tightening response. The three panels of each variable share the vertical scale.}
\end{figure}

\begin{figure}[ht]
    \centering
    \includegraphics[width=\linewidth]{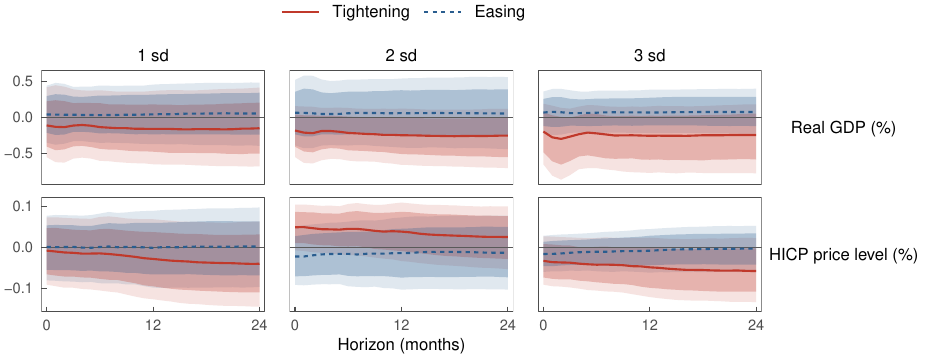}
    \caption{Impulse response functions of output and the price level to monetary policy shocks of different sign and size.}\label{fig:main-irfs_macro}\vspace*{-0.25cm}
    \caption*{\footnotesize \textit{Notes}: Real GDP in percent, HICP as the cumulative price-level response in percent. Columns: absolute shock sizes of $1$, $2$ and $3$ sd; rows: variables, with a shared vertical scale within each row. Each panel pairs tightening and easing. Median, credible sets and normalization as in Figure \ref{fig:main-irfs_credit}. Industrial production, unemployment and the financial variables: Appendix \ref{app:addresults}.}
\end{figure}

Figures \ref{fig:main-irfs_credit} and \ref{fig:main-irfs_macro} show the dynamic responses behind these peaks for the credit factors and for the two headline aggregates, output and the price level; the financial variables and unemployment are covered in Appendix \ref{app:addresults}. Each panel pairs tightening and easing shocks of the same absolute size for one variable, showing their posterior medians and $50$ and $68$ percent credible sets, normalized per $1$ sd of tightening so that a conventionally signed easing response has the same sign as the tightening response.

Figure \ref{fig:main-irfs_credit} shows the responses of the credit-condition factors. Following a $1$ sd contractionary shock, the credit standards, bank lending and borrower balance sheet factors tighten gradually and persistently, with median responses that build up over the entire horizon and peak around $0.35$, $0.30$ and $0.30$ standard deviations toward the end of the second year; credit demand declines in the median. Under expansionary shocks none of the credit factors displays a response whose band excludes zero at any horizon, for any of the three easing sizes. The normalized medians are close to zero during the first year.

Figure \ref{fig:main-irfs_macro} contains the responses of output and the price level. The median output response is negative throughout, with a scaled peak of roughly $-0.17$ percent for a $1$ sd and $-0.30$ percent for a $3$ sd shock, notably smaller than the large-shock industrial production response. The price level declines in the median after a $1$ sd tightening, with a scaled peak of roughly $-0.04$ percent that materializes only toward the end of the two-year horizon ($-0.06$ percent for $3$ sds). For $2$ sds a wrong-signed response emerges, which can be traced back to the impact estimate.

Turning to easing, apart from the impact response of the 3-month Euribor (Section \ref{sec:impactasym}), no variable shows a conventionally signed response whose band excludes zero at any horizon, for any of the three easing sizes. Large ($3$ sd) easings do produce two responses whose bands exclude zero in the baseline --- but strikingly, both have the wrong sign. The stock market falls on impact (scaled impact of about $0.9$ in the flipped normalization, i.e., prices decline by about $2.7$ percent after a $3$ sd easing, visible on the easing side of the stock price panel of Figure \ref{fig:peaksize}), and the corporate spread widens during the first half year (Appendix \ref{app:addresults}). Large easing shocks are rare and concentrated in crisis episodes and months (Section \ref{sec:data}). These include September $2001$, October $2008$, the euro area crisis months of $2011$ and November $2020$, in which stock prices fell, spreads widened and unemployment rose for reasons unrelated to policy. Our conjecture is that the corresponding estimates are informed by these months, which might explain why the wrong-signed results emerge.\footnote{This discussion is again related to the results of \citet{kolesar2024dynamic} and \citet{gouletcoulombe2025blackbox}. The average marginal effects would mask such heterogeneity (the region below $-2$ sds receives $12$ percent of the weight); the nonparametric model provides an explicit estimate, and the ``wrong-signed'' response is the price of characterizing the nonlinearity rather than summarizing it.}

\inlinehead{Robustness and alternative model specifications} All results so far refer to the baseline specification, a single-regime model with an active outlier component of Section \ref{sec:econometrics}. Our robustness specification grid varies two dimensions, the treatment of the error variances (outlier-robust versus fully homoskedastic) and the regime configuration (a single regime versus two regimes defined by recessions, the ELB or a positive credit-to-GDP gap). The two-regime models estimate all regime-specific functions on the respective subsamples; we discuss these results in more detail in Section \ref{sec:heterogeneities}.

The substantive robustness question is the treatment of extreme observations, because the sample includes the pandemic. Comparing the outlier-robust and homoskedastic full-sample models, we find the following. The credit factors' medians are essentially unaffected; the homoskedastic variant is somewhat more precise, with bands of the credit standards, borrower balance sheet and bank lending factors excluding zero in a few scenarios where the baseline bands include it. Among the real and financial variables, the qualitative results survive. The tightening responses of stock prices, the corporate spread and unemployment, and the flat easing side, appear in both variants. The one substantive divergence is the magnitude of the industrial production response to a large tightening, whose band excludes zero at nearly all horizons in both variants but which is almost three times larger in the baseline ($-0.76$ versus $-0.28$), so the sign of this result is robust while its size is not. The remaining divergences concern output, the price level and the large-easing cells, where the homoskedastic variant produces somewhat tighter bands around similar or smaller medians and more wrong-signed easing responses with bands excluding zero. Specification sensitivity thus concentrates in the responses that were puzzling to begin with, which supports the conjecture that a few unusual months drive them; that the homoskedastic model produces more such responses suggests that the outlier scaling absorbs this partially. The corresponding figures are in Appendix \ref{app:empirical}.

\subsection{Heterogeneities across time}\label{sec:heterogeneities}
We now examine whether monetary policy effects vary over time. Our framework distinguishes two conceptually different (although related) sources of such variation. The first operates through {initial conditions}. The second operates through the {transmission mechanism}. Figure \ref{fig:shockts} shows that both positive and negative shocks occur in every regime, so the results below are not driven by particular signs of shocks clustering in particular regimes.

\begin{figure}[ht]
    \centering
    \includegraphics[width=\linewidth]{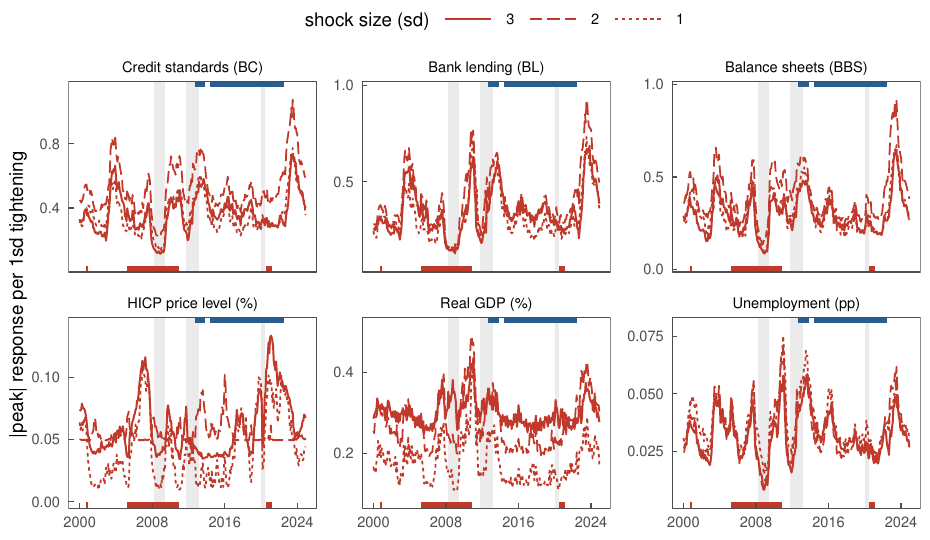}
    \caption{Median peak responses to tightening shocks over time, computed conditionally on the initial conditions of each month in the sample.}
    \label{fig:tbyt}\vspace*{-0.25cm}
    \caption*{\footnotesize \textit{Notes}: Absolute peak over $24$ months of the median response to a $3$ sd (solid), $2$ sd (long-dashed) and $1$ sd (dotted) tightening, normalized per $1$ sd, for each month's initial conditions. Regime markers as in Figure \ref{fig:shockts}: gray shading, recessions; blue bar (top), ELB months; red bar (bottom), positive credit-to-GDP gap.}
\end{figure}

\inlinehead{Time variation through initial conditions} We generally compute the GIRFs of Equation (\ref{eq:definitionIRF}) separately for each month in the sample, conditioning on the respective initial conditions, whereas the previous results were time averages of these granular estimates. Because the impact response within a regime does not depend on $\bm{z}_t$, any time variation materializes at propagation horizons. Figure \ref{fig:tbyt} summarizes the resulting peak responses over time for the tightening direction and six selected variables; the easing counterpart is in Appendix \ref{app:addresults}. 

For the price level, the median (across time) peak scaled response to a large tightening is around $0.05$ percent, but reaches $0.13$ for shocks occurring around $2021$; for output, peaks reach $0.43$ for shocks occurring in early $2011$; and for the credit standards, bank lending and borrower balance sheet factors, peak responses reach $0.63$ to $0.73$ for shocks occurring in $2023$, roughly twice their typical size. These comparisons concern posterior medians. The $68$ percent sets of responses at individual dates are wide --- for the credit standards factor in mid-$2023$ roughly $[0.05, 1.65]$ around a median of $0.73$, against a typical median of $0.35$ --- so the variation over time is suggestive rather than sharply estimated. The episodes of particularly strong transmission do not line up cleanly with the recession, ELB or credit-gap windows, which suggests that the specific configuration of initial conditions rather than these regimes drives transmission strength.\footnote{These patterns are broadly consistent with the concurrent work of \citet{scalone2026endogenous}, who use a smooth-transition local projection framework for the euro area in which the state, the change in the private-sector debt service ratio, responds endogenously to past shocks. In their estimates, tightening raises debt-servicing burdens and amplifies subsequent tightenings. Some of the largest peak responses in Figure \ref{fig:tbyt} arise for shocks around 2023 and 2024, after the fastest tightening cycle in our sample.} For the credit factors, the peak-response paths of small and large tightenings track each other closely, so the time variation in the credit responses is driven by the initial conditions rather than by the shock size; for output, the large-shock path lies persistently above the small-shock path, and for the price level it does so at most dates.

\inlinehead{Regime-based state dependence} Given this variation in medians, one might expect the predefined regime splits to capture at least part of it. They essentially do not. In the context of the initial conditions, we compare responses of the full-sample model averaged over regime versus non-regime initial conditions, with the averages formed within posterior draws. Across all three regime classifications, all six shock sizes, both variance specifications and all $14$ variables, the $68$ percent bands of the regime and non-regime responses never separate for more than two isolated horizons. Note that these comparisons come with a caveat. The regime subsamples differ considerably in size ($39$ months of recession, $113$ months at the ELB and $81$ months with a positive credit-to-GDP gap), and the recession regime in particular is thin, so the absence of band separation may reflect limited information in the regime subsample rather than genuinely identical transmission.

\begin{figure}[ht]
    \centering
    \includegraphics[width=\linewidth]{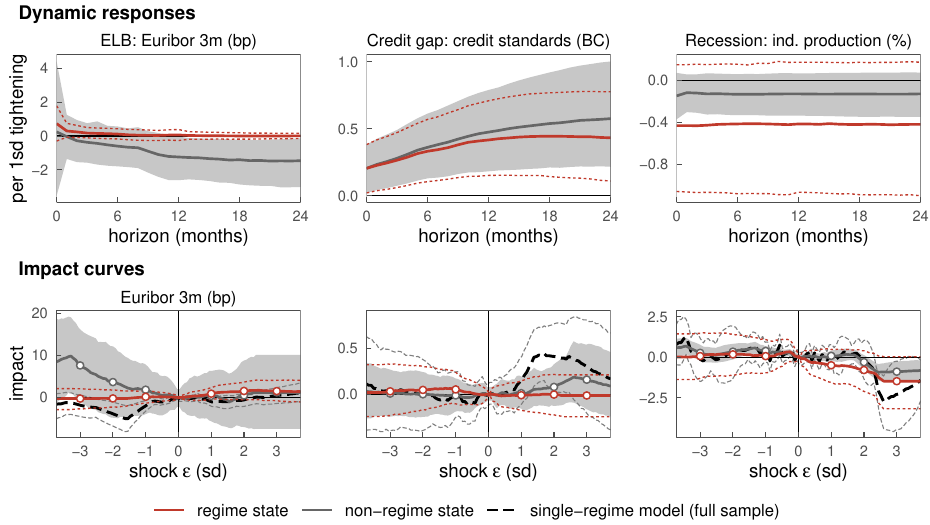}
    \caption{Regime contrasts under tightening: dynamic responses ($2$ sd, normalized) and regime-specific impact curves.}
    \label{fig:statedep}\vspace*{-0.25cm}
    \caption*{\footnotesize \textit{Notes}: Top: responses to a $2$ sd tightening, normalized per $1$ sd; gray: non-regime median and $68$ percent band; red: regime median with the edges of its $68$ percent band (dotted). ELB and recession columns: two-regime models; credit-gap column: single-regime model averaged over positive- versus nonpositive-gap initial conditions. Bottom: impact curves of the respective two-regime model for the variable of the column, with the single-regime curve overlaid (black dashed; thin gray: its $68$ percent set). Circles: median impact at $\varepsilon \in \{\pm1,\pm2,\pm3\}$.}
\end{figure}

Comparing responses from the regime-specific models ($S=2$) across their two regimes, the bands are distinct in exactly one case, which is the response of the 3-month Euribor at the ELB (left column of Figure \ref{fig:statedep}). This separation emerges only for larger tightenings. The bands are separate at longer horizons (from about one and a half years) with $2$ and $3$ sd shocks, whereas at $1$ sd and for easings of any size they overlap at every horizon (Appendix \ref{app:addresults}). Figure \ref{fig:statedep} shows the $2$ sd case, for which each state of every two-regime model still contains at least two tightenings of that size. Outside the ELB, the short rate reverses below baseline about a year after a tightening, while at the bound the response stays at zero. The model imposes no bound on the simulated short rate, so this difference is not mechanical; it is consistent with constrained short-rate adjustment at the ELB being reflected in the regime-specific functions. The other two columns of Figure \ref{fig:statedep} are representative of the null result. They show the credit standards factor across credit-gap states (middle column, from the single-regime model; discussed below) and industrial production across recessions and expansions from the recession model (right column), where the regime response is directionally stronger but the sets of the two states overlap at every horizon.

On impact, state dependence operates only through the regime-specific estimates. The bottom row of Figure \ref{fig:statedep} compares the subsample-estimated regime curves with the single-regime curve, and Appendix \ref{app:addresults} shows the credit standards factor, the corporate spread, industrial production and the price level for all three regime models. The regime curves are mostly flatter than the single-regime curve, and their credible sets overlap across states in every comparison. The corporate spread keeps a flat easing side and a rising tightening side in most regime and non-regime curves but is V-shaped in the positive-gap state, whereas the credit standards factor responds to neither sign in the regime state of any of the three models. The credit-gap contrast also reveals sensitivity to model specification. In the full (non-regime) model, tightening produces weaker median credit responses on positive-gap dates but the credible sets overlap at every horizon. The credit-gap regime model instead yields near-zero median credit-factor responses in both states. The flatter regime curves, and the near-zero credit-factor responses in the positive-gap state of the credit-gap regime model, partly reflect the limited number of tail observations in the regime states, as discussed in Section \ref{sec:data}.

\section{Conclusions}\label{sec:conclusions}
We explore whether the effects of monetary policy in the euro area depend on the sign and size of the policy shock and on the state of the economy. We combine monthly macro-financial data with quarterly survey-based measures of bank lending conditions in a mixed-frequency model estimated with BART. Neither the impact nor the propagation of the shock is restricted to a prespecified functional form. Our findings are threefold. First, sign asymmetry is the dominant nonlinearity. Contractionary shocks lower stock prices, widen spreads, raise unemployment and tighten credit conditions, whereas expansionary shocks of any size produce mostly insignificant results. Second, the survey-based credit factors tighten gradually and persistently after contractionary shocks, a dimension of transmission that the muted response of the short-term interest rate to tightenings misses. Third, credible sets do not separate across recessions, the ELB or credit-to-GDP gap states, apart from the short-term rate at the ELB, yet median responses vary considerably with the initial conditions from which a shock occurs, albeit with wide credible sets at individual dates.

From a policy perspective, the results suggest that contractionary policy works rather reliably and tightens bank lending conditions persistently, whereas easing shocks of any size do not loosen them significantly. Effective easing thus might require complementary tools such as unconventional policy measures. The configuration of macro-financial initial conditions when a shock occurs seems to matter more for its transmission than coarse regimes. From a modeling perspective, the lesson for empirical and theoretical work is that state dependence alone does not seem to be enough. Empirical specifications that allow transmission to vary across predefined regimes but impose symmetry within them would miss the patterns documented here. Likewise, theoretical models that generate state dependence from a regime switch but respond symmetrically to shocks within a state have a hard time rationalizing these types of asymmetry.

{\setstretch{1.15}\putbib}\normalsize
\clearpage

\end{bibunit}

\clearpage\doublespacing\normalsize
\appendix
\thispagestyle{empty}
\part*{\singlespacing\centering\sffamily\LARGE Online Appendix}
\begin{center}
\vspace*{-1cm}\singlespacing{\LARGE\sffamily\textbf{\titletext}}\\[1em]
\end{center}

\setcounter{page}{1}
\setcounter{section}{0}
\setcounter{equation}{0}
\setcounter{footnote}{0}
\setcounter{figure}{0}
\setcounter{table}{0}
\renewcommand\thesection{\Alph{section}}
\renewcommand\theequation{\Alph{section}.\arabic{equation}}
\renewcommand\thefigure{\Alph{section}.\arabic{figure}}
\renewcommand\thetable{\Alph{section}.\arabic{table}}
\makeatletter\@addtoreset{equation}{section}\@addtoreset{figure}{section}\@addtoreset{table}{section}\makeatother

\vspace*{0.5cm}
\appendixtableofcontents\clearpage

\begin{bibunit}
\appsection{Model and computation}\label{app:technical}
\appsubsection{Simulation of the GIRFs}\label{app:girf}
Section \ref{sec:GIRFdescription} defines the GIRF of variable $i$ at horizon $h$ as the difference between two conditional expectations, Equation (\ref{eq:definitionIRF}), in which the period-$t$ shock is set to $d_0 + d$ or to the baseline $d_0$, and the expectations integrate over the shocks and reduced-form innovations of periods $t,\hdots,t+h$ given the initial conditions $\bm{z}_t$ and the regime $s_t = s$. The baseline is fixed at the sample mean of the standardized shock over the effective estimation sample, which equals zero up to the initialization months dropped for the lags, so that we refer to it as $d_0 = 0$. The baseline path thus carries no period-$t$ shock and the shocked path adds $d$.\footnote{Other choices in the literature are the observed shock or averaging over a grid or random draws of $d_0$.} This appendix describes how the expectations are evaluated.

Let $k$ index the posterior draws of the sampler described in Appendix \ref{app:inference}, and let $r$ index the origin periods of the effective estimation sample, $r = 1,\hdots,T$, each of which serves as one history. For a given draw $k$ and origin $r$, the experiment is defined by the following recursion.
\begin{enumerate}
    \item \textit{Initial state.} The initial conditions are the $p$ months preceding the shock period, $\bm{z}_r^{(k)} = (\bm{y}_{r-1}^{(k)\prime},\hdots,\bm{y}_{r-p}^{(k)\prime})'$, taken from the completed data of draw $k$. The monthly variables are observed, and the latent monthly counterparts of the quarterly variables are the draws of the latent states. Because the latent states are redrawn in every iteration, $\bm{z}_r^{(k)}$ varies across $k$. The regime is that of the origin, $s = s_{r}$, and it is held fixed at all horizons. The regime-specific functions $\bm{H}_s^{(k)}$ and $\bm{G}_s^{(k)}$ and the covariance $\bm{\Sigma}_s^{(k)}$ of the origin regime are used along the entire simulated path, so regimes do not transition during the simulation.
    \item \textit{Impact.} The impact is the first simulated period. Its shock is replaced by $d_0 + d$ on the shocked path and by $d_0$ on the baseline path, and the period is simulated from Equation (\ref{eq:MFBART}), $\bm{y}_{r}^{(c,k)} = \bm{H}_s^{(k)}(\bm{z}_r^{(k)}) + \bm{G}_s^{(k)}(c) + \bm{u}_{r}^{(k)}$ for $c \in \{d_0 + d, d_0\}$, with the same innovation $\bm{u}_r^{(k)}$ on both paths. The impact response of the experiment is therefore $\bm{G}_s^{(k)}(d_0 + d) - \bm{G}_s^{(k)}(d_0)$, which does not depend on the initial conditions.
    \item \textit{Propagation.} For $h = 1,2,\hdots$, both paths are simulated forward, $\bm{y}_{r+h}^{(c,k)} = \bm{H}_s^{(k)}(\bm{z}_{r+h}^{(c,k)}) + \bm{G}_s^{(k)}(\varepsilon_{r+h}^{(k)}) + \bm{u}_{r+h}^{(k)}$, where $\bm{z}_{r+h}^{(c,k)}$ stacks the $p$ most recent simulated (and, for $h < p$, initial) values of the respective path. The future shocks $\varepsilon_{r+h}^{(k)}$ are drawn independently across horizons with replacement from the set of realized shocks of the effective estimation sample, because we do not assume a distribution for the shock; the innovations $\bm{u}_{r+h}^{(k)}$ are drawn from $\mathcal{N}(\bm{0}_n,\bm{\Sigma}_s^{(k)})$; and the outlier scalings along the simulated paths are set to $o_{i,r+h} = 1$, so the counterfactuals reflect regular-variance periods rather than extrapolated outlier states. No restrictions are imposed on the simulated paths beyond the value of the period-$r$ shock.
    \item \textit{Common random numbers.} The shocked and the baseline path share the same draws of future shocks and innovations. For each history and posterior draw we simulate $30$ pairs of paths, and the response for history $r$, $\hat{\bm{\delta}}_{r,h}^{(d;k)}$, is the difference between the average shocked and the baseline path across these simulations. Sharing the random inputs reduces the simulation noise in this difference; because the two paths pass through the nonlinear functions from different states, the noise does not cancel exactly at $h \geq 1$. See also \citet{pfarrhofer2025scenario}.
    \item \textit{Averaging.} The reported response for a set of histories $\mathcal{R}$ (the full sample, or the months of a regime) is the average over the histories within each draw, $\hat{\bm{\delta}}_{h}^{(d;k)} = |\mathcal{R}|^{-1}\sum_{r\in\mathcal{R}}\hat{\bm{\delta}}_{r,h}^{(d;k)}$, and quantiles are taken across draws $k$. The credible bands therefore reflect posterior uncertainty about the averaged response.
\end{enumerate}

\appsubsection{Parametric reference forms for sign and size asymmetries}\label{app:funcforms}
Less flexible but more easily interpretable measures of asymmetries can be obtained by assuming a specific functional form for the shock function of each equation $i = 1,\hdots,n$,
\begin{equation*}
    g_i(\varepsilon_t) = {\beta}_{i1} \varepsilon_t + {\beta}_{i2} \tilde{g}(\varepsilon_t).
\end{equation*}
This is the approach discussed in, e.g., \citet{caravello2024disentangling}. A linear contemporaneous response is obtained by setting $\beta_{i2} = 0$; the impact response is then $d \cdot \beta_{i1}$, constant over time and proportional in $d$, and nonlinearities can only arise at propagation horizons through $\bm{H}(\bullet)$. For sign nonlinearities, $\tilde{g}(\varepsilon_t) = |\varepsilon_t|$, so the function has a kink at zero with easing-side slope $\beta_{i1} - \beta_{i2}$ and tightening-side slope $\beta_{i1} + \beta_{i2}$; for size nonlinearities, $\tilde{g}(\varepsilon_t) = \varepsilon_t|\varepsilon_t|$, which produces signed curvature. The impact response for baseline $d_0$ is $d \cdot \beta_{i1} + (\tilde{g}(d_0 + d) - \tilde{g}(d_0))\beta_{i2}$. These three forms are the reference forms fitted to every posterior draw of the nonparametric impact function in Section \ref{sec:impactasym}, and serve as descriptive approximations of the unrestricted $g_{si}$.

\appsubsection{Bayesian inference}\label{app:inference}
\inlinehead{Tree priors}
The conditional mean functions are approximated equation-by-equation and regime-by-regime by sums of $B$ regression trees, as described in Section \ref{sec:bart}. Due to the large number of trees, regularization is essential in BART. It can be achieved via priors on the tree-generating stochastic process, which contains the prior assumptions about the tree structures $\mathcal{T}_{si,b}$. A prior on node depth determines the probability of a node being nonterminal, which decreases with depth. The probability of a node at depth $\mathfrak{d}$ being nonterminal is set to $\alpha/(1+\mathfrak{d})^\beta$, where $\alpha \in (0,1)$ and $\beta \in \mathbb{R}^+$. For each regime, we use default values, $\alpha=0.95$ and $\beta=2$, as specified by \citet{chipman2010bart}, which have been shown to work well in many contexts. Further, we specify a discrete uniform prior over the splitting variables, which implies that all variables are equally likely to act as determining the partitions of the input space. Lastly, the thresholds within the splitting rules are also assigned a uniform prior over the range of the relevant splitting variables.

We now turn to the prior parameters of the terminal nodes. With $\#\text{TN}_{si,b}$ denoting the number of terminal node parameters of tree $b$, we choose independent Gaussian priors that are symmetric across trees for the terminal node parameters, such that $\mu_{si,bl}\sim \mathcal{N}(0,v_{si})$ for $l=1,\hdots,\#\text{TN}_{si,b}$. The prior variance is based on $\sqrt{v_{si}} = [\max_{t\in \mathcal{S}_s}(y_{it}) - \min_{t\in \mathcal{S}_s}(y_{it})] / (2\gamma \sqrt{B})$, where $\mathcal{S}_s$ is the set of observations in regime $s$ and we set $\gamma = 1.96$. This implies that most prior mass is put on the observed range of values (by regime). The prior on tree depth favors simple rather than complex trees, and the tightness of the prior on the terminal node parameters increases with the number of trees. We choose $B=250$ for the number of trees, which strikes a balance between having too few trees and using needlessly many of them.

\inlinehead{Other specification details}
As stated in Equation (\ref{eq:MFBART}), the reduced-form errors feature a regime-specific covariance matrix $\bm{\Sigma}_s$ alongside variable-specific outlier scalings collected in $\bm{O}_t = \diag(o_{1t},\hdots,o_{nt})$, so that the effective error covariance in period $t$ is $\bm{O}_t\bm{\Sigma}_{s_t}\bm{O}_t$. These scalings may capture huge-variance shocks and outliers such as during the pandemic of the early $2020$s. Tree-based conditional means do not extrapolate linearly from extreme predictor values and can capture some forms of heteroskedasticity without an explicit treatment in the error terms \citep[see][for a discussion]{clark2021tail}. To provide further robustness and reflect the recent literature \citep[see, e.g.,][]{lenza2022estimate}, we introduce an additional safeguard against such observations. 

Specifically, we use a variant of \citet{carriero2021addressing}, and assume for each equation $i = 1,\hdots,n$:
\begin{equation*}
    o_{it} = 1 \text{ with probability } 1 - \mathfrak{p}_i, \quad o_{it} \sim \mathcal{U}(2,6) \text{ with probability } \mathfrak{p}_i,
\end{equation*}
where $\mathcal{U}(\bullet)$ is a discrete uniform distribution with support between $2$ and $6$ and $\mathfrak{p}_i$ is the equation-specific probability of observing an outlier. Each $\mathfrak{p}_i$ is assigned a $\mathcal{B}(1, 99)$ prior --- implying an expected prior outlier frequency of one percent --- and is updated from its conditional Beta posterior given the current outlier indicators of equation $i$.

The regime-specific matrices $\bm{\Sigma}_s$ are assigned the hierarchical inverse Wishart prior of \citet{huang2013simple}, as suggested by \citet{esser2024seemingly}. Under this hierarchy, the implied marginal prior on each error sd $\sqrt{\Sigma_{s,ii}}$ is half-t with $\nu = 2$ degrees of freedom and scale $A_i$. We calibrate $A_i$ variable by variable such that $75$ percent of the implied prior mass on the sd lies below the corresponding OLS residual sd (with the resulting scales clipped to $[0.1, 10]$), anchoring the prior to the scale of each series without imposing a dogmatic value. In the homoskedastic variant of our specification grid (see Appendix \ref{app:specgrid}), the outlier component is switched off, i.e., $o_{it} = 1$ for all $i$ and $t$; comparing the two variants serves as a robustness check on the influence of extreme observations.

\inlinehead{Approximate sampling of the latent states}
{\sloppy Let $\bm{x}_t = (\bm{z}_t',\varepsilon_t)'$ collect lags and the shock, and write the conditional mean of Equation (\ref{eq:MFBART}) in regime $s$ as $\bm{F}_s(\bm{x}_t) = \bm{H}_s(\bm{z}_t) + \bm{G}_s(\varepsilon_t)$. Define $\bm{y} = (\bm{y}_1',\hdots,\bm{y}_T')'$, ${\bm{F}} = \left(\bm{F}_{s_1}({\bm{x}}_1)',\hdots,\bm{F}_{s_T}({\bm{x}}_T)'\right)'$, and $\bm{\epsilon} = (\bm{\epsilon}_1',\hdots,\bm{\epsilon}_T')'$, and let $\bm{X} = ({\bm{x}}_1,\hdots,{\bm{x}}_T)'$. Then we may rewrite the state equation in stacked notation as:\par}
\begin{equation*}
    \bm{y} = {\bm{F}} + \bm{\epsilon}, \quad \bm{\epsilon}_t \sim \mathcal{N}(\bm{0}_{n},\bm{O}_t\bm{\Sigma}_{s_t}\bm{O}_t),
\end{equation*}
with the errors independent across time. Conditional on the trees, $\bm{F}$ is a nonlinear function of the latent states, which precludes exact filtering. Following the strategy employed in a multivariate mixed-frequency BART context by \citet{huber2023nowcasting} \citep[inspired by][]{ish2019interpreting,crawford2018bayesian}, we work with a linear approximation $\bm{y} = \bm{X}{\bm{A}} + \tilde{\bm{\epsilon}}$, where ${\bm{A}} = ({\bm{A}}_1,\hdots,{\bm{A}}_p)'$ are linearized dynamic coefficients, partitioned such that ${\bm{A}}_j$ is associated with the $j$th lag in ${\bm{x}}_t$; see also \citet{marcellino2025nonparametric} for a more technical discussion. Rather than obtaining $\bm{A}$ from a pseudo-inverse of $\bm{X}$, we estimate it equation by equation with a Bayesian regression of the current BART fit on the predictors, $\bm{f}_i = \bm{X}\bm{a}_i + \bm{e}_i$ with $\bm{e}_i \sim \mathcal{N}(\bm{0}_T, \sigma_{\text{apx},i}^2\bm{I}_T)$, where $\bm{f}_i$ stacks the fitted values of the $i$th conditional mean function. The coefficients $\bm{a}_i$ carry a horseshoe prior \citep{horseshoe}, sampled via the auxiliary representation of \citet{makalic2015simple}, and the approximation-error variance $\sigma_{\text{apx},i}^2$ has an inverse Gamma $\mathcal{IG}(100, 0.01)$ prior --- tight around small values --- and is updated in closed form. This delivers both a shrunken linearization and an explicit estimate of its accuracy.

The linearized system is a standard linear Gaussian state space model, and we rely on the precision sampler of \citet{chan2009efficient}, which renders estimation of large systems feasible. The estimated approximation-error variances are propagated into the filter. The period-$t$ error covariance used when sampling the latent states is $\bm{O}_t\bm{\Sigma}_{s_t}\bm{O}_t + \diag(\sigma_{\text{apx},1}^2,\hdots,\sigma_{\text{apx},n}^2)$, so that the uncertainty about the latent monthly processes reflects the quality of the linearization. In the regime specifications ($S > 1$), the linear approximation is computed on the full sample and shared across regimes, which stabilizes the filter where regime subsamples are short. The conditional mean functions, shock functions and error covariances remain regime-specific in their specification; the latent states drawn under the shared linearization enter the subsequent draws of these regime-specific objects, as they do in the single-regime model.

\inlinehead{Sampling algorithm}
Our Gibbs sampler alternates between the model parameters and the latent states (the unobserved high-frequency processes behind the low-frequency variables). Conditional on the latent states, we (i) update the regression trees equation-by-equation and regime-by-regime, with separate ensembles for $h_{si}(\bm{z}_t)$ and $g_{si}(\varepsilon_t)$, as weighted regressions that account for the cross-equation error correlation and the outlier scalings (see \citet{esser2024seemingly} and \citet{pfarrhofer2025scenario} for the posterior moments); (ii) draw the regime-specific covariance matrices $\bm{\Sigma}_s$ from their inverse Wishart posteriors on the standardized residuals, followed by the hierarchical scales of the \citet{huang2013simple} prior; and (iii) draw the outlier scalings $o_{it}$ from their discrete closed-form conditional posteriors and the outlier probabilities $\mathfrak{p}_i$ from their Beta posteriors. Conditional on all parameters, the latent states are drawn jointly with a precision sampler based on the linearized dynamic coefficients of the approximation described above. After each draw of the latent states, the completed data matrix is re-standardized to mean zero and unit variance before the next parameter step. Because each latent draw is approximately standardized in any case, this rescaling is an approximation that stabilizes the sampler. 

We run this algorithm for $12{,}000$ iterations, discarding the initial $3{,}000$ draws as burn-in and retaining every third of the remaining draws. These draws are used to compute the impact curves and the dynamic causal effects described in Appendix \ref{app:girf}.

\FloatBarrier
\appsection{Data, specifications and robustness}\label{app:empirical}
\appsubsection{Data}
Table \ref{tab:appdata_VAR} lists all variables of the system together with their transformations and sources. The data span January $1999$ to December $2024$; with $p = 12$ lags, the effective estimation sample covers January $2000$ to December $2024$, that is, $T = 300$ high-frequency observations. Real GDP is observed for $104$ quarters ($1999$Q$1$ to $2024$Q$4$), while the survey-based series are available for $86$ quarters ($2003$Q$1$ to $2024$Q$2$); the mixed-frequency setup handles the ragged availability of the quarterly series without discarding monthly information, and the latent monthly credit conditions of the $42$ months of the effective sample outside the survey coverage (January $2000$ to December $2002$ and July to December $2024$) are inferred by the model. The macro-financial information set coincides roughly with recent related work, see, for instance, \citet{jarocinski2020deconstructing}.

\begin{table*}[ht]
\caption{Data, transformations and source.}\label{tab:appdata_VAR}\vspace*{-1.5em}
\begin{center}
\begin{threeparttable}
\scriptsize
\begin{tabular*}{\textwidth}{@{\extracolsep{\fill}} llll}
\toprule
\textbf{Variable} & \textbf{Description} & \textbf{FQ / $h(x)$} & \textbf{Source} \\
\midrule
Credit \texttt{(BC)} & BLS factor: broad credit channel & Q / 1 & ECB BLS, own calc. \\
Lending \texttt{(BL)} & BLS factor: bank lending channel & Q / 1 & ECB BLS, own calc. \\
Balance sheet \texttt{(BBS)} & BLS factor: borrower balance sheet channel & Q / 1 & ECB BLS, own calc. \\
Demand \texttt{(D)} & BLS factor: credit demand & Q / 1 & ECB BLS, own calc. \\
Output \texttt{(rGDP)} & Real gross domestic product, s.a. & Q / 4 & ECB Data Portal \\
Production \texttt{(IP)} & Industrial production, w.a., s.a. & M / 3 & ECB Data Portal \\
Unemployment \texttt{(UNEMP)} & Unemployment rate, s.a. & M / 5 & ECB Data Portal \\
Prices \texttt{(HICP)} & Harmonized Index of Consumer Prices, s.a. & M / 4 & ECB Data Portal \\
Short rate \texttt{(Euribor3m)} & 3-month Euribor & M / 1 & ECB Data Portal \\
1-year rate \texttt{(Euribor1y)} & 1-year Euribor & M / 1 & ECB Data Portal \\
Long rate \texttt{(LTR10y)} & 10-year long-term interest rate & M / 1 & ECB Data Portal \\
Stock market \texttt{(ES50)} & Euro Stoxx 50 equity index & M / 2 & ECB Data Portal \\
Spreads \texttt{(OAS)} & ICE BofA Euro High Yield Index OAS & M / 1 & FRED \\
Exchange rate \texttt{(EER)} & Nominal effective exchange rate & M / 2 & ECB Data Portal \\
\midrule
MP shock \texttt{(MPPC)} & High-frequency monetary policy shock & M / 1 & EA-EMPD, own calc. \\
Credit-to-GDP gap & Regime indicator: positive gap & Q / 1 & ECB Data Portal \\
\bottomrule
\end{tabular*}
\begin{tablenotes}[para,flushleft]
\scriptsize{\textit{Notes}: Sources: FRED (\href{https://fred.stlouisfed.org/}{fred.stlouisfed.org}) and the ECB Data Portal (\href{https://data.ecb.europa.eu/}{data.ecb.europa.eu}). BLS factors: first principal components of the blocks in Table \ref{tab:appdata_BLS}; monetary policy shock: EA-EMPD \citep{altavilla2025monetary}, Appendix \ref{app:datadetails}; the credit-to-GDP gap defines the credit-gap regime (positive gap), carried forward from quarters to months. FQ: frequency; OAS: option-adjusted spread; w.a./s.a.: working day/seasonally adjusted. Transformations $h(x_t)$: (1) $x_t$; (2) $100\log x_t$; (3) $100\log(x_t/x_{t-1})$; (4) $f \cdot 100\log(x_t/x_{t-1})$, $f$ observations per year; (5) $\Delta x_t$. Responses of differenced series are cumulated and, for (4), divided by the annualization factor $f$ of the observed series; the HICP response is the cumulative price-level response. Impact curves (Section \ref{sec:impactasym}, Appendix \ref{app:gxcurves}) are in the same units per sd of the shock, except basis points for interest rates and the spread.}
\end{tablenotes}
\end{threeparttable}
\end{center}
\end{table*}

\begin{table*}[ht]
\caption{BLS items and corresponding transmission channel.}\label{tab:appdata_BLS}
\centering
\begin{threeparttable}
\footnotesize
\begin{tabular}{l | l}
\toprule
\textbf{BLS Variable Description} & \textbf{Channel} \\
\midrule
\multicolumn{2}{l}{Change in credit standards for $\dots$} \\
\midrule
$\dots$ loans to enterprises & \multirow{3}{*}{(Broad) Credit} \\
$\dots$ loans to HHs for house purchase &  \\
$\dots$ consumer credit and other lending to HHs &  \\
\midrule
\multicolumn{2}{l}{Change in demand for $\dots$} \\
\midrule
$\dots$ loans to enterprises & \multirow{3}{*}{Demand} \\
$\dots$ loans to HHs for house purchase &  \\
$\dots$ consumer credit and other lending to HHs &  \\
\midrule
\multicolumn{2}{l}{Factors contributing to changes in credit standards for loans to enterprises} \\
\midrule
Costs related to bank's capital position & \multirow{6}{*}{Bank lending} \\
Bank's ability to access market financing &  \\
Bank's liquidity position &  \\
Competition from other banks &  \\
Competition from non-banks &  \\
Competition from market financing &  \\
\midrule
General economic situation and outlook & \multirow{3}{*}{Borrower balance sheet} \\
Industry or firm-specific situation and outlook &  \\
Risk related to the collateral demanded &  \\
\midrule
\multicolumn{2}{l}{Factors contributing to changes in credit standards for loans for house purchase} \\
\midrule
Costs related to bank's capital position & \multirow{6}{*}{Bank lending} \\
Cost of funds and balance sheet constraints (until 2021Q4) &  \\
Bank's ability to access market financing &  \\
Bank's liquidity position &  \\
Competition from other banks &  \\
Competition from non-banks &  \\
\midrule
General economic situation and outlook & \multirow{3}{*}{Borrower balance sheet} \\
Housing market prospects, incl. expected house price developments &  \\
Borrower's creditworthiness &  \\
\midrule
\multicolumn{2}{l}{Factors contributing to changes in credit standards for consumer credit and other lending} \\
\midrule
Costs related to bank's capital position & \multirow{6}{*}{Bank lending} \\
Cost of funds and balance sheet constraints (until 2021Q4) &  \\
Bank's ability to access market financing &  \\
Bank's liquidity position &  \\
Competition from other banks &  \\
Competition from non-banks &  \\
\midrule
General economic situation and outlook & \multirow{2}{*}{Borrower balance sheet} \\
Creditworthiness of consumers &  \\
\bottomrule
\end{tabular}
\begin{tablenotes}[para,flushleft]
\scriptsize{\textit{Notes}: Quarterly euro area net percentages over the past three months (tightened minus eased; for demand, increased minus decreased), ECB BLS statistics. ``Cost of funds and balance sheet constraints'' was discontinued with the $2022$ questionnaire revision (Appendix \ref{app:datadetails}).}
\end{tablenotes}
\end{threeparttable}
\end{table*}

\FloatBarrier
\appsubsection{Construction of the BLS factors and the monetary policy shock}\label{app:datadetails}
\inlinehead{BLS factors}
The BLS collects information on bank lending conditions in the euro area.\footnote{See \href{https://data.ecb.europa.eu/methodology/bank-lending-survey-bls}{data.ecb.europa.eu/methodology/bank-lending-survey-bls} for a detailed description of the BLS.} The survey is addressed to senior loan officers of a representative sample of euro area banks and is conducted on a quarterly basis. We follow \citet{ciccarelli2013heterogeneous} in using the information from BLS questions that describe changes in credit standards for and demand for loans to firms and households (for house purchase as well as consumer credit and other lending), as well as factors that affected these changes in lending conditions over the past three months. In particular, we use information from the question on changes in credit standards to define the broad \textit{credit} factor and banks' answers about changes in loan demand to characterize credit \textit{demand}. Furthermore, we use information from questions about the factors that contributed to changes in credit standards relating to the bank \textit{lending} channel and the \textit{borrower balance sheet} channel; Table \ref{tab:appdata_BLS} maps the individual survey items to these blocks.

The input series are the euro area aggregate net percentages published by the ECB, defined as the difference between the share of banks reporting that credit standards (or the contributing factor) have contributed to tightening and the share reporting an easing contribution, or, for demand, the share reporting an increase minus the share reporting a decrease. The series are the ECB Data Portal BLS series with aggregation code \texttt{WFNET}, that is, plain net percentages (not the response-intensity-weighted diffusion index) with the euro area aggregate weighted by each country's share in outstanding loans. Items that are not available over the full sample are retained as series with missing values, which the imputation step below fills. The item ``cost of funds and balance sheet constraints'' was discontinued with the $2022$ questionnaire revision, several bank-lending items for household loans are only available from $2022$Q$2$, and borrower creditworthiness for house purchase from $2015$Q$2$. The panel was downloaded in July $2024$ and covers $2003$Q$1$ to $2024$Q$2$.

For each of the four blocks in Table \ref{tab:appdata_BLS}, we extract one common factor from the quarterly items of the block with the iterative principal-component procedure of \citet{stock2002macroeconomic}, which also handles missing values: (i) each item is standardized to zero mean and unit variance using its available observations, and missing values are initially replaced by zero, the mean of the standardized series; (ii) the first principal component is computed; (iii) missing values are replaced by the common component (factor times loadings) of the current estimate, the completed panel is re-standardized, and step (ii) is repeated; the iterations stop when the change in the mean squared factor between iterations falls below $10^{-6}$ or after $50$ iterations. Observed values are never altered by the procedure; only missing values are filled. The resulting factor is a linear combination of the standardized items, so it is measured in standardized survey units rather than in net percentages. 

The routine normalizes the loadings rather than the factor, so the factors do not have unit variance; we divide their responses by the sample standard deviation of the respective factor throughout, so that a reported response of, e.g., $0.5$ means that the factor moves by half of its typical variation over the sample. The sign of a principal component is arbitrary and not normalized by the routine; we flip all four factors so that they load positively on every item of their block, which orients them like the underlying net percentages: an increase indicates tighter credit standards (or, for demand, higher demand). The factors enter the model without further transformation.

\inlinehead{Monetary policy shock}
The shock enters the model as observed and exogenous data, which simplifies estimation and the computation of the dynamic causal effects (Section \ref{sec:econometrics_mfvar}). We construct it from the high-frequency surprises in the EA-EMPD \citep{altavilla2025monetary}, which records intraday changes in OIS rates and stock prices around Governing Council decisions and around speeches by the ECB President and Executive Board members. The construction proceeds in five steps.
\begin{enumerate}
    \item \textit{Event sample.} We use the Governing Council monetary event windows (database code \texttt{GC\_ME}), the speeches of Executive Board members (\texttt{EB}) and of the President (\texttt{P}), and treat every event as one observation. The database version we use runs from January $1999$ to October $2025$.
    \item \textit{Surprise variables.} For each event we take the surprises in the OIS rates at maturities of 1, 2, 3 and 6 months and 1 year, and in the Euro Stoxx 50. Each of the six series is standardized.
    \item \textit{Interest rate factor.} One common factor is extracted from the five standardized OIS surprises across all events with the same iterative principal-component procedure as for the BLS factors, which also fills missing quotes at individual maturities. This yields a single event-level score of the yield-curve surprise.
    \item \textit{Sign restrictions.} Following \citet{jarocinski2020deconstructing}, we consider the bivariate system of the interest rate factor and the stock price surprise, using the events for which both are available. Let $\bm{L}$ be the lower Cholesky factor of their covariance matrix across events. We draw $10{,}000$ orthogonal matrices $\bm{Q}$ and, for each, form the candidate impact matrix $\bm{B}_0^{-1} = \bm{L}\bm{Q}$ of Section \ref{sec:data}, normalized to have a positive diagonal. A candidate is accepted if the first shock $\varepsilon_\tau$ raises the interest rate factor $s_{\tau}^{\text{(OIS)}}$ and lowers the stock price surprise $s_{\tau}^{\text{(ES50)}}$ (a monetary policy shock) and the second shock $\tilde{\varepsilon}_\tau$ raises both (a central bank information shock); draws that do not satisfy the restrictions are redrawn. For each accepted matrix, the structural shocks of every event are obtained as $(\varepsilon_{\tau}, \tilde{\varepsilon}_{\tau})' = \bm{B}_0 (s_{\tau}^{\text{(OIS)}}, s_{\tau}^{\text{(ES50)}})'$, and the monetary policy shock of an event is the median of $\varepsilon_\tau$ across the $10{,}000$ accepted rotations.
    \item \textit{Aggregation.} The event-level shocks are summed within each calendar month. Before estimation, the monthly series is regressed (without intercept) on its first lag and replaced by the residual, which removes first-order autocorrelation, and then standardized to zero mean and unit variance over the full data span (January $1999$ to December $2024$). A positive value is a tightening shock.
\end{enumerate}
To express the shock in basis points, we divide the covariance of $\varepsilon_t$ with the monthly sum of the raw 3-month (1-year) OIS surprises by the variance of $\varepsilon_t$ over the effective sample, which gives $2.70$ ($2.64$) basis points per unit of the shock (Section \ref{sec:data}).

Two remarks are in order. First, high-frequency surprises can exhibit predictability and residual central bank information content \citep[see, e.g.,][]{nakamura2018high,miranda2021transmission,bauer2022reassessment}; the sign-restriction separation and projecting out autocorrelation are designed to mitigate these concerns. Second, the interest rate factor and the rotation-based separation are estimated inputs that enter the model as data. By using the loadings and the median across accepted rotations, we condition on point estimates of the shock series, so uncertainty from the factor extraction and the identification step is not propagated into the credible sets we report --- a simplification often used in the broader shocks-as-data literature.

\begin{figure}[ht]
    \centering
    \includegraphics[width=\linewidth]{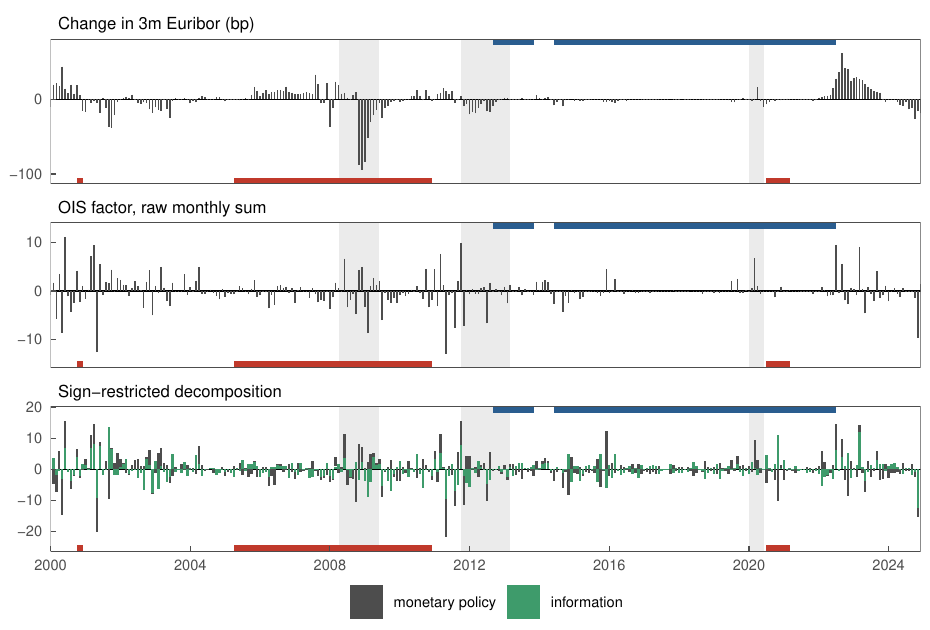}
    \caption{From the raw OIS surprises to the monetary policy shock.}
    \label{fig:shock_decomp}\vspace*{-0.25cm}
    \caption*{\footnotesize \textit{Notes}: Monthly, January $2000$ to December $2024$. Top: change in the 3-month Euribor (basis points). Middle: monthly sum of the event-level OIS factor scores (standardized event units, no adjustment). Bottom: monetary policy (gray) and information (green) components of the sign-restricted decomposition, stacked; medians across rotations, summed within the month. Shading and bars as in Figure \ref{fig:shockts}.}
\end{figure}

Figure \ref{fig:shock_decomp} places the raw surprises next to actual interest rate changes. It shows the monthly change in the 3-month Euribor, the monthly sum of the event-level OIS factor scores before any adjustment, and the monetary policy and information components of the sign-restricted decomposition.

\appsubsection{Model specifications}\label{app:specgrid}
The specification grid varies along two dimensions. The first is the treatment of the error variances. The baseline features the outlier component described in Appendix \ref{app:inference} (\texttt{outl}), while a homoskedastic variant (\texttt{homo}) switches this component off and serves as a robustness check --- particularly relevant because the pandemic observations are part of our sample. The second is the regime configuration. The single-regime model ($S = 1$, \texttt{full}) is complemented by three two-regime models ($S = 2$) that split the sample into recessions versus expansions (\texttt{rec}), periods at versus away from the ELB (\texttt{elb}), and months with a positive versus a nonpositive credit-to-GDP gap (\texttt{fin}). In the regime models, both the conditional mean functions and the error covariance matrix are regime-specific.

\begin{figure}[ht]
    \centering
    \includegraphics[width=\linewidth]{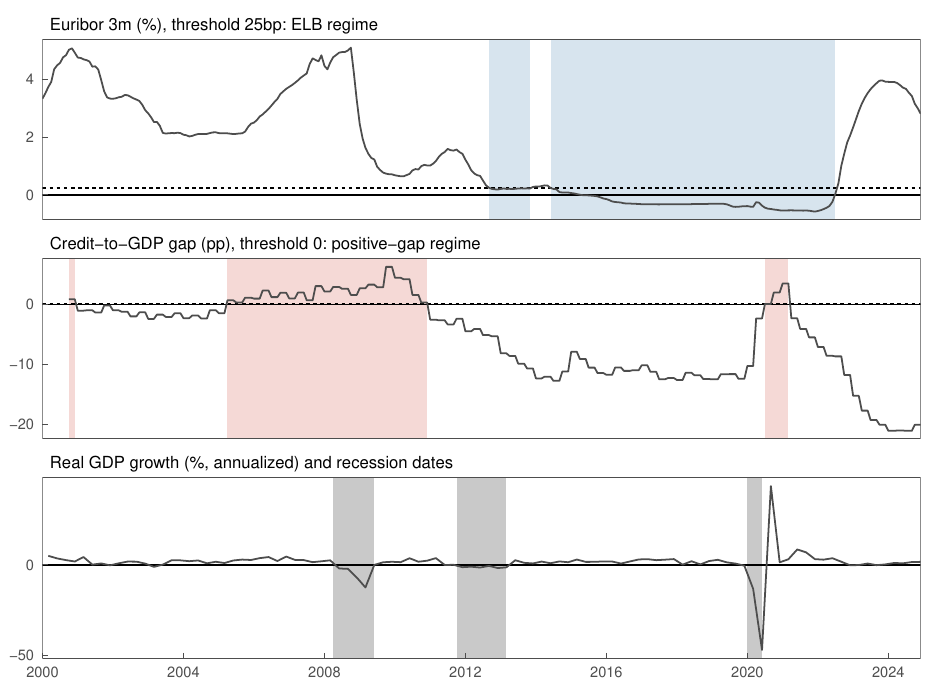}
    \caption{Regime-splitting variables, thresholds and regime months.}
    \label{fig:regime_vars}\vspace*{-0.25cm}
    \caption*{\footnotesize \textit{Notes}: Top: 3-month Euribor (percent), dotted line at $25$ basis points, blue shading: ELB months. Middle: credit-to-GDP gap (percentage points), threshold zero, red shading: positive-gap months. Bottom: annualized quarterly real GDP growth (percent), gray shading: recessions. Shaded months are the regime dates used in estimation and in Figure \ref{fig:shockts}.}
\end{figure}

Figure \ref{fig:regime_vars} shows the variables underlying the three regime classifications together with the thresholds and the resulting regime months, which are the ones marked in Figure \ref{fig:shockts}.

\begin{figure}[ht]
    \centering
    \includegraphics[width=\linewidth]{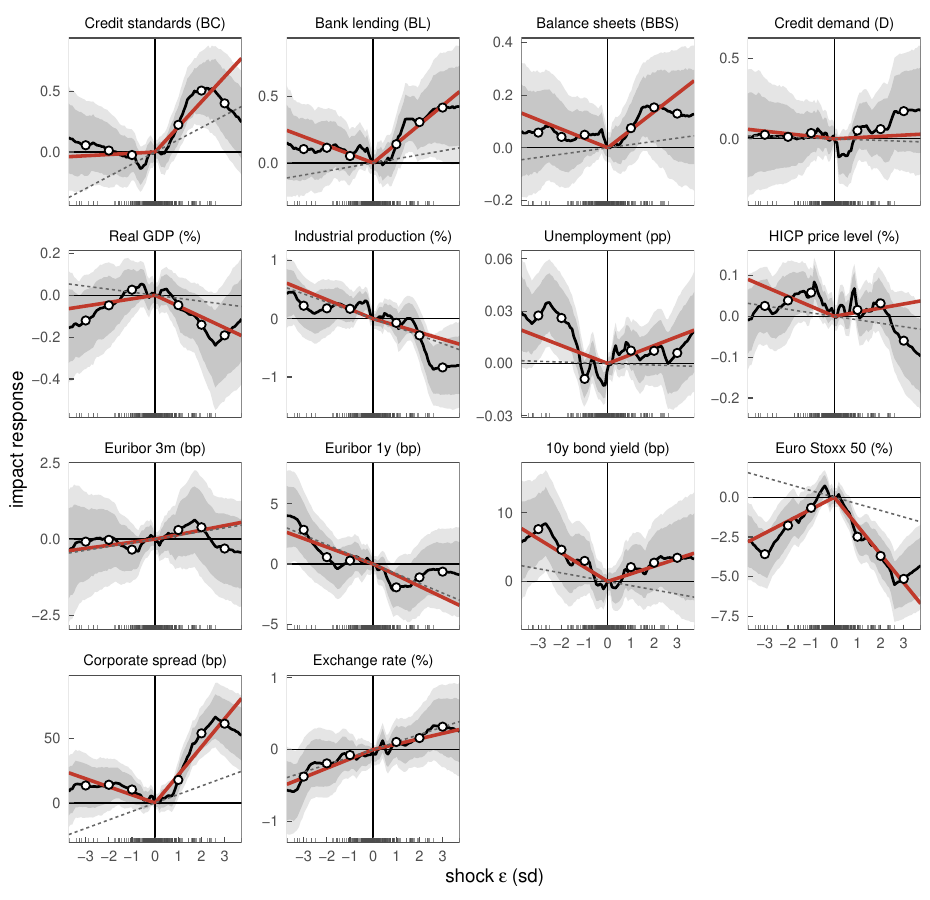}
    \caption{Impact curves for the full system: homoskedastic variant.}
    \label{fig:gx_app_homo}\vspace*{-0.25cm}
    \caption*{\footnotesize \textit{Notes}: Homoskedastic single-regime model (\texttt{homo}/\texttt{full}); layout as in Figure \ref{fig:shock_vars}: median with $50/68$ percent sets (black, gray), kink and linear fits at the realized shocks (red; dashed gray), median impacts at $\varepsilon \in \{\pm1,\pm2,\pm3\}$ (circles), realized shocks (rug). Units as in Figures \ref{fig:gx_app_bls}--\ref{fig:gx_app_financial}.}
\end{figure}

\begin{figure}[p]
    \centering
    \includegraphics[width=\linewidth]{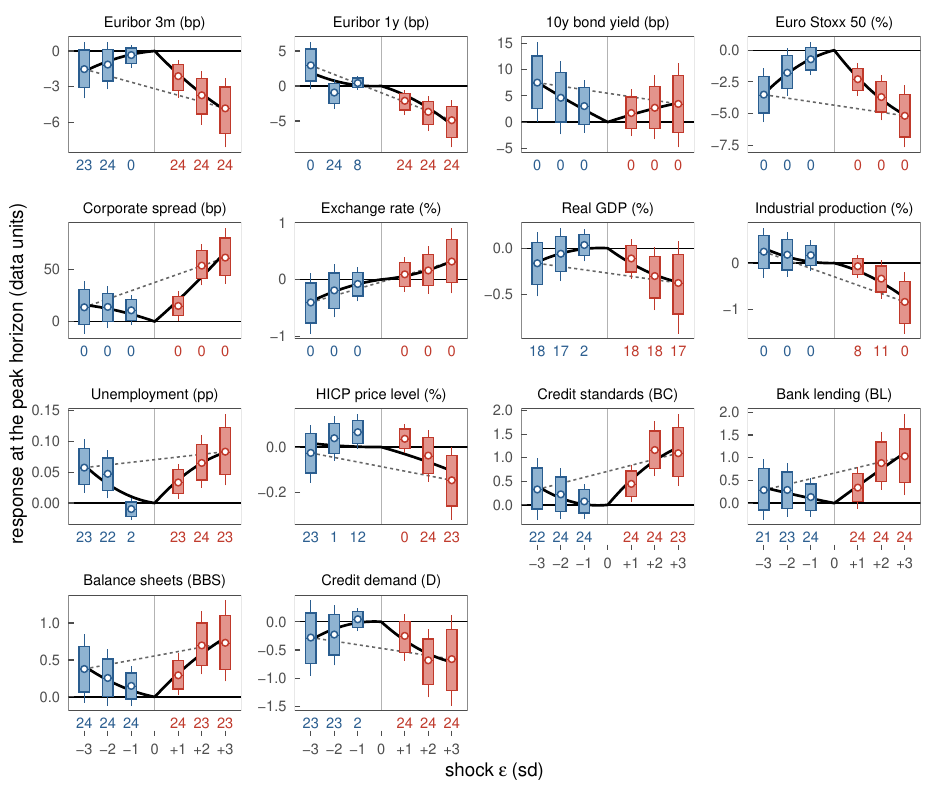}
    \caption{Peak responses as a function of the size and sign of the shock: homoskedastic variant.}
    \label{fig:app_peaksize_homo}\vspace*{-0.25cm}
    \caption*{\footnotesize \textit{Notes}: As Figure \ref{fig:peaksize} for the homoskedastic single-regime model (\texttt{homo}/\texttt{full}).}
\end{figure}

\begin{figure}[ht]
    \centering
    \includegraphics[width=\linewidth]{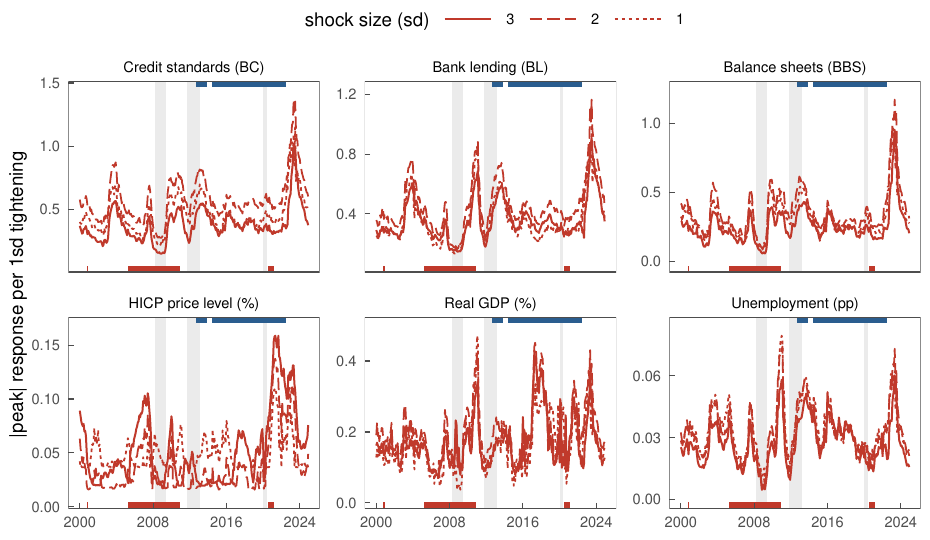}
    \caption{Peak responses to tightening shocks over time: homoskedastic variant.}
    \label{fig:app_tbyt_homo}\vspace*{-0.25cm}
    \caption*{\footnotesize \textit{Notes}: As Figure \ref{fig:tbyt} for the homoskedastic single-regime model (\texttt{homo}/\texttt{full}). Medians only.}
\end{figure}

\begin{figure}[ht]
    \centering
    \includegraphics[width=\linewidth]{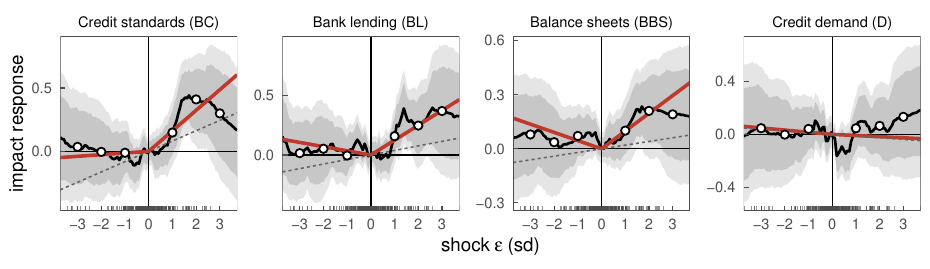}
    \caption{Impact curves: survey-based credit-condition factors.}
    \label{fig:gx_app_bls}\vspace*{-0.25cm}
    \caption*{\footnotesize \textit{Notes}: Baseline specification; layout as in Figure \ref{fig:shock_vars}: median with $50/68$ percent sets (black, gray), kink and linear fits at the realized shocks (red; dashed gray), median impacts at $\varepsilon \in \{\pm1,\pm2,\pm3\}$ (circles), realized shocks (rug).}
\end{figure}

\begin{figure}[ht]
    \centering
    \includegraphics[width=\linewidth]{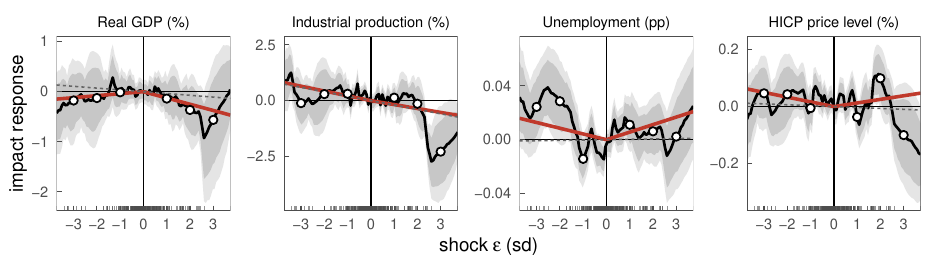}
    \caption{Impact curves: real activity and prices.}
    \label{fig:gx_app_real}\vspace*{-0.25cm}
    \caption*{\footnotesize \textit{Notes}: As Figure \ref{fig:gx_app_bls}; real GDP and HICP repeat the panels of Figure \ref{fig:shock_vars}.}
\end{figure}

\begin{figure}[ht]
    \centering
    \includegraphics[width=\linewidth]{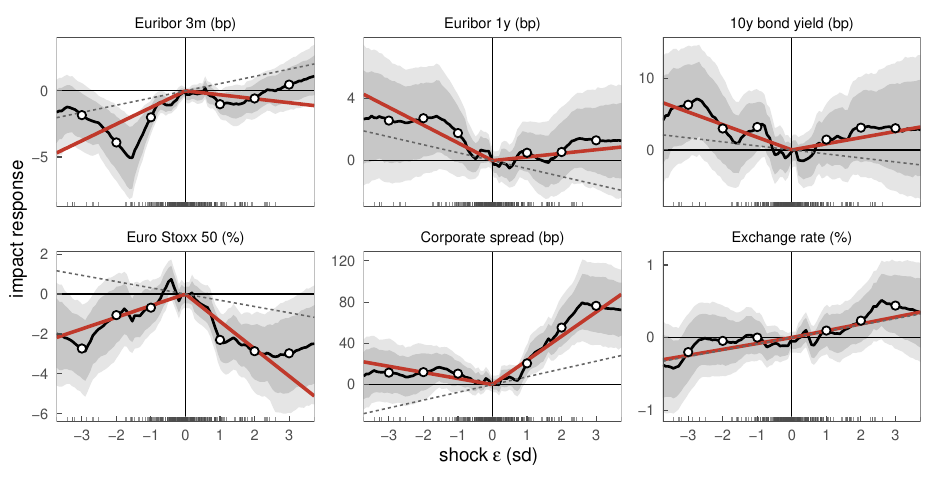}
    \caption{Impact curves: financial variables.}
    \label{fig:gx_app_financial}\vspace*{-0.25cm}
    \caption*{\footnotesize \textit{Notes}: As Figure \ref{fig:gx_app_bls}.}
\end{figure}

\appsubsection{Robustness details}\label{app:robustness}
This section collects the figures underlying the robustness discussion in Section \ref{sec:dynamics}. Figure \ref{fig:gx_app_homo} shows the impact curves of the homoskedastic variant for all variables of the system, the counterpart of Figures \ref{fig:gx_app_bls}--\ref{fig:gx_app_financial}; Figure \ref{fig:app_peaksize_homo} shows its peak responses by size and sign of the shock, the counterpart of Figure \ref{fig:peaksize}; and Figure \ref{fig:app_tbyt_homo} shows its peak responses over time for tightening shocks, the counterpart of Figure \ref{fig:tbyt}. The two-regime models are discussed in Section \ref{sec:heterogeneities}; regime versus non-regime impacts are in Appendix \ref{app:stateimpact}.

\FloatBarrier
\appsection{Supplementary results}\label{app:addresults}

\appsubsection{Impact curves for the full system}\label{app:gxcurves}
Figures \ref{fig:gx_app_bls}--\ref{fig:gx_app_financial} show the estimated impact curves for all variables of the system, complementing the selected panels in Figure \ref{fig:shock_vars} and the parametric approximations in Table \ref{tab:gxfits_full}.

\begin{figure}[ht]
    \centering
    \includegraphics[width=\linewidth]{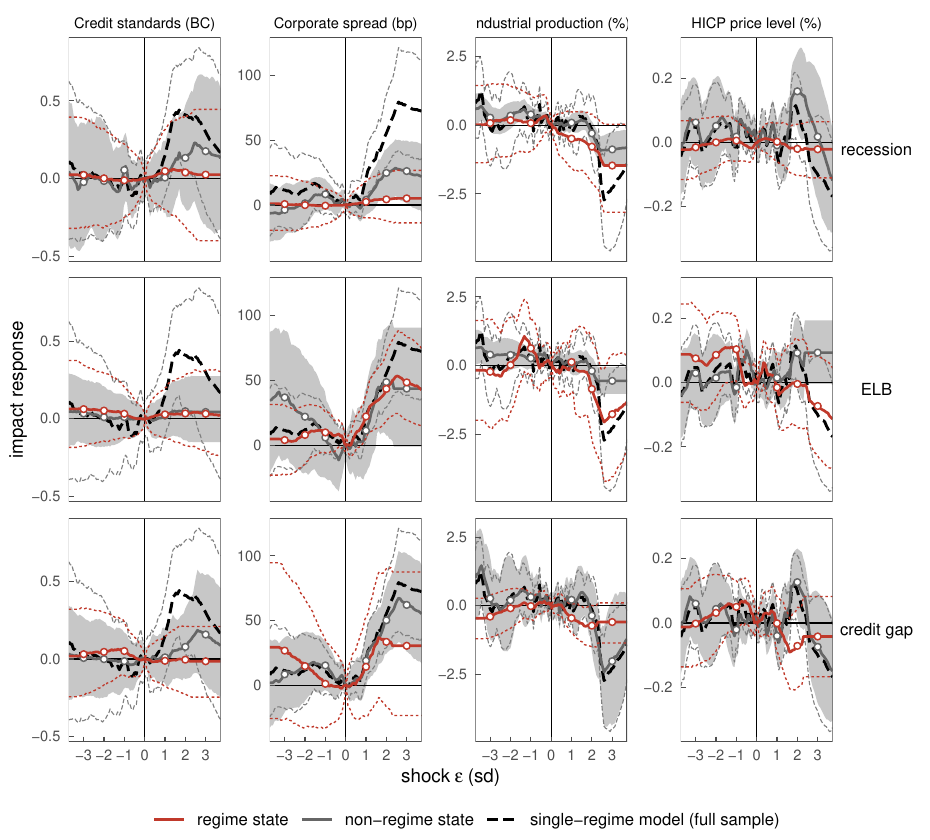}
    \caption{Regime versus non-regime impact curves of the two-regime models.}
    \label{fig:gx_state}\vspace*{-0.25cm}
    \caption*{\footnotesize \textit{Notes}: Rows: two-regime models. Lines, bands and circles as in the bottom row of Figure \ref{fig:statedep}.}
\end{figure}

\FloatBarrier
\appsubsection{Impact curves across regimes}\label{app:stateimpact}
This section complements the discussion of state dependence at impact in Section \ref{sec:heterogeneities}. Figure \ref{fig:gx_state} compares the regime-specific impact curves, with the full-sample curve of the single-regime model overlaid for reference, for four representative variables: the credit standards factor, the corporate spread, industrial production and the HICP. The figure also allows assessing whether the shape of the impact nonlinearity documented in Section \ref{sec:impactasym} differs across regimes.

\begin{figure}[ht]
    \centering
    \includegraphics[width=\linewidth]{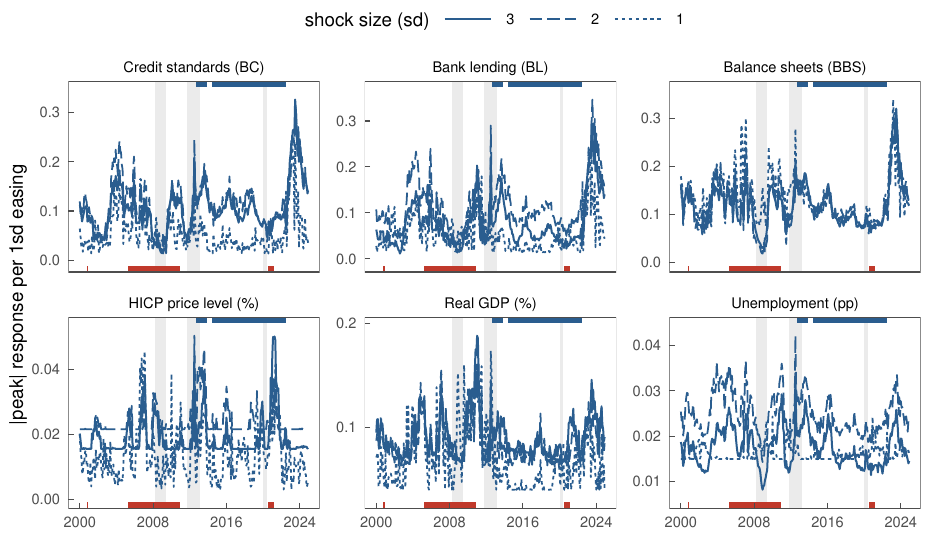}
    \caption{Peak responses to easing shocks over time, conditional on the initial conditions of each month in the sample.}
    \label{fig:app_tbyt_easing}\vspace*{-0.25cm}
    \caption*{\footnotesize \textit{Notes}: As Figure \ref{fig:tbyt} for easing shocks, normalized per $1$ sd of tightening. Medians only.}
\end{figure}

\FloatBarrier
\appsubsection{Time variation of peak responses}\label{app:tbyt}
Figure \ref{fig:tbyt} in the paper shows the peak responses over time for tightening shocks discussed in Section \ref{sec:heterogeneities}; Figure \ref{fig:app_tbyt_easing} shows the easing counterpart. Both figures display posterior medians only.

\FloatBarrier
\appsubsection{Complete scenario sets}\label{app:fullsets}
Several figures in the paper display a subset of the computed objects for readability. This section provides the complementary sets.

\begin{figure}[!htbp]
    \centering
    \includegraphics[width=\linewidth]{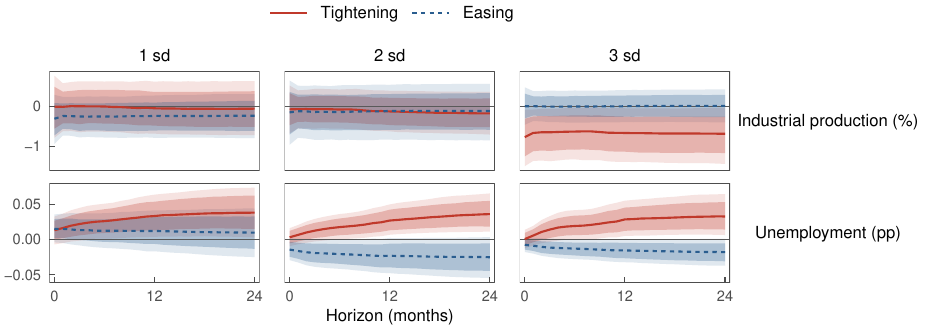}
    \caption{Impulse responses of industrial production and unemployment.}
    \label{fig:app_real_fan}
\par
    \includegraphics[width=\linewidth]{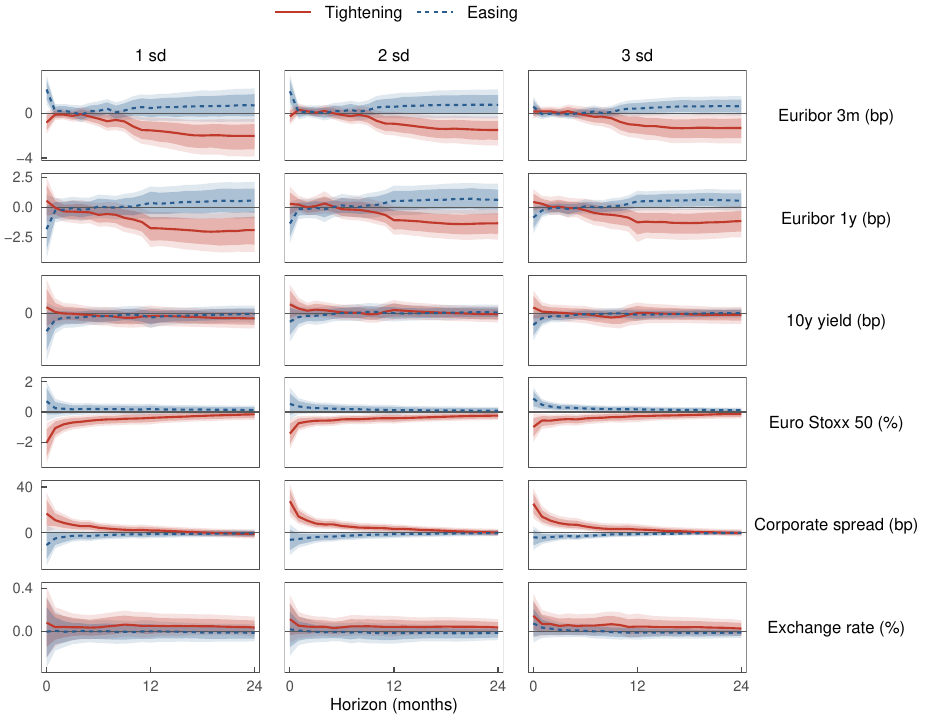}
    \caption{Impulse responses of the financial variables.}
    \label{fig:app_fin_fan}\vspace*{-0.25cm}
    \caption*{\footnotesize \textit{Notes}: Both figures: medians, credible sets and normalization as in Figure \ref{fig:main-irfs_credit}; columns: absolute shock sizes; rows: variables.}
\end{figure}

\begin{figure}[ht]
    \centering
    \includegraphics[width=\linewidth]{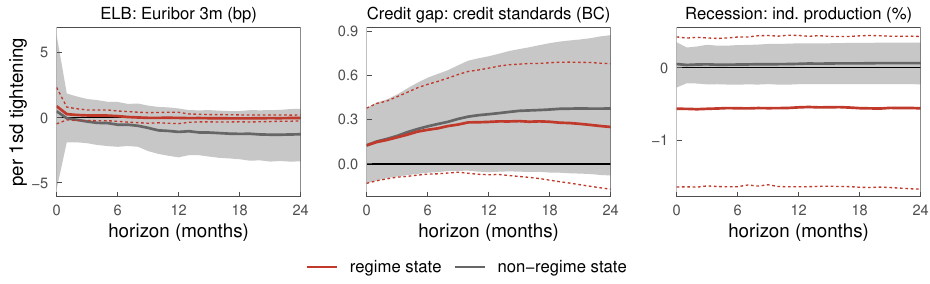}
    \caption{The three regime contrasts under a $1$ sd tightening shock.}
    \label{fig:app_statedep_1sd}\vspace*{-0.25cm}
    \caption*{\footnotesize \textit{Notes}: As the top row of Figure \ref{fig:statedep}, for a $1$ sd instead of a $2$ sd tightening.}
\end{figure}

\begin{figure}[ht]
    \centering
    \includegraphics[width=\linewidth]{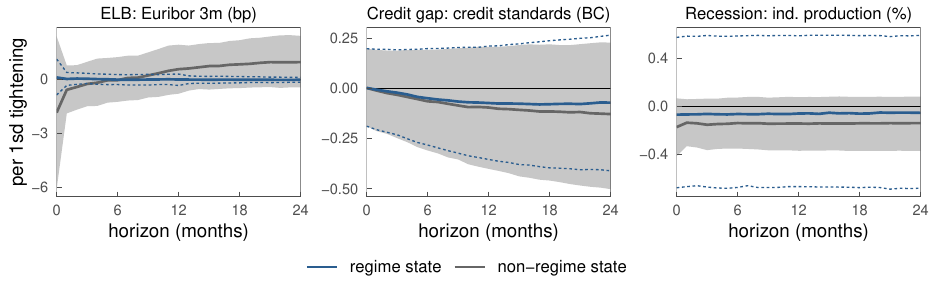}
    \caption{The three regime contrasts under a $-2$ sd easing shock.}
    \label{fig:app_statedep_easing}\vspace*{-0.25cm}
    \caption*{\footnotesize \textit{Notes}: As the top row of Figure \ref{fig:statedep}, with the regime state in blue.}
\end{figure}

\FloatBarrier\clearpage
{\setstretch{1.2}\putbib}\normalsize
\end{bibunit}

\end{document}

%% file: plots_paper/tab_gx_fits.tex
\begin{tabular}{@{}llr@{}lr@{}lr@{}lr@{}lrrr@{}}
\toprule
 & & \multicolumn{2}{c}{Linear form} & \multicolumn{6}{c}{Sign form $\beta_1\varepsilon + \beta_2|\varepsilon|$} & \multicolumn{3}{c}{BIC winner share (\%)} \\
\cmidrule(lr){3-4}\cmidrule(lr){5-10}\cmidrule(lr){11-13}
Variable & Unit & \multicolumn{2}{c}{\shortstack{Slope\\$\beta_1$}} & \multicolumn{2}{c}{\shortstack{Easing\\$\beta_1 - \beta_2$}} & \multicolumn{2}{c}{\shortstack{Tightening\\$\beta_1 + \beta_2$}} & \multicolumn{2}{c}{\shortstack{Difference\\$2\beta_2$}} & \multicolumn{1}{c}{Linear} & \multicolumn{1}{c}{Sign} & \multicolumn{1}{c}{Size} \\
\midrule
\multicolumn{13}{l}{\emph{Financial}}\\
Euribor 3m & bp & $[\phantom{-}0.1,$ & $\phantom{-}1.0]^{*}$ & $[\phantom{-}0.5,$ & $\phantom{-}2.1]^{*}$ & $[-0.9,$ & $\phantom{-}0.2]^{\phantom{*}}$ & $[-2.7,$ & $-0.6]^{*}$ & 1 & 84 & 16 \\
Euribor 1y & bp & $[-1.1,$ & $\phantom{-}0.1]^{\phantom{*}}$ & $[-2.6,$ & $\phantom{-}0.3]^{\phantom{*}}$ & $[-0.9,$ & $\phantom{-}1.4]^{\phantom{*}}$ & $[-0.9,$ & $\phantom{-}3.8]^{\phantom{*}}$ & 3 & 77 & 20 \\
10y bond yield & bp & $[-1.8,$ & $\phantom{-}0.7]^{\phantom{*}}$ & $[-4.2,$ & $\phantom{-}0.6]^{\phantom{*}}$ & $[-1.9,$ & $\phantom{-}3.8]^{\phantom{*}}$ & $[-1.7,$ & $\phantom{-}7.2]^{\phantom{*}}$ & 5 & 72 & 22 \\
Corporate spread & bp & $[\phantom{-}3.6,$ & $\phantom{-}11.6]^{*}$ & $[-15.1,$ & $\phantom{-}1.9]^{\phantom{*}}$ & $[\phantom{-}14.5,$ & $\phantom{-}33.0]^{*}$ & $[\phantom{-}14.0,$ & $\phantom{-}45.9]^{*}$ & 1 & 95 & 4 \\
Euro Stoxx 50 & \% & $[-0.628,$ & $-0.022]^{*}$ & $[-0.013,$ & $\phantom{-}1.185]^{\phantom{*}}$ & $[-2.120,$ & $-0.664]^{*}$ & $[-3.188,$ & $-0.786]^{*}$ & 1 & 84 & 14 \\
Exchange rate & \% & $[\phantom{-}0.005,$ & $\phantom{-}0.173]^{*}$ & $[-0.085,$ & $\phantom{-}0.253]^{\phantom{*}}$ & $[-0.083,$ & $\phantom{-}0.274]^{\phantom{*}}$ & $[-0.291,$ & $\phantom{-}0.317]^{\phantom{*}}$ & 5 & 69 & 26 \\
\addlinespace[0.5em]
\multicolumn{13}{l}{\emph{Real activity and prices}}\\
Real GDP & \% & $[-0.104,$ & $\phantom{-}0.029]^{\phantom{*}}$ & $[-0.112,$ & $\phantom{-}0.190]^{\phantom{*}}$ & $[-0.300,$ & $\phantom{-}0.041]^{\phantom{*}}$ & $[-0.457,$ & $\phantom{-}0.128]^{\phantom{*}}$ & 9 & 71 & 20 \\
Industrial production & \% & $[-0.290,$ & $-0.105]^{*}$ & $[-0.439,$ & $\phantom{-}0.006]^{\phantom{*}}$ & $[-0.424,$ & $\phantom{-}0.064]^{\phantom{*}}$ & $[-0.389,$ & $\phantom{-}0.471]^{\phantom{*}}$ & 16 & 65 & 19 \\
Unemployment & pp & $[-0.004,$ & $\phantom{-}0.004]^{\phantom{*}}$ & $[-0.012,$ & $\phantom{-}0.004]^{\phantom{*}}$ & $[-0.004,$ & $\phantom{-}0.015]^{\phantom{*}}$ & $[-0.006,$ & $\phantom{-}0.026]^{\phantom{*}}$ & 3 & 56 & 41 \\
HICP & \% & $[-0.015,$ & $\phantom{-}0.009]^{\phantom{*}}$ & $[-0.046,$ & $\phantom{-}0.013]^{\phantom{*}}$ & $[-0.020,$ & $\phantom{-}0.045]^{\phantom{*}}$ & $[-0.028,$ & $\phantom{-}0.086]^{\phantom{*}}$ & 9 & 72 & 19 \\
\addlinespace[0.5em]
\multicolumn{13}{l}{\emph{Credit (BLS factors)}}\\
Credit standards (BC) & sd & $[\phantom{-}0.014,$ & $\phantom{-}0.148]^{*}$ & $[-0.112,$ & $\phantom{-}0.136]^{\phantom{*}}$ & $[\phantom{-}0.032,$ & $\phantom{-}0.290]^{*}$ & $[-0.062,$ & $\phantom{-}0.361]^{\phantom{*}}$ & 5 & 68 & 26 \\
Bank lending (BL) & sd & $[-0.012,$ & $\phantom{-}0.088]^{\phantom{*}}$ & $[-0.132,$ & $\phantom{-}0.065]^{\phantom{*}}$ & $[\phantom{-}0.015,$ & $\phantom{-}0.234]^{*}$ & $[-0.029,$ & $\phantom{-}0.338]^{\phantom{*}}$ & 6 & 77 & 17 \\
Balance sheets (BBS) & sd & $[-0.024,$ & $\phantom{-}0.068]^{\phantom{*}}$ & $[-0.132,$ & $\phantom{-}0.043]^{\phantom{*}}$ & $[\phantom{-}0.002,$ & $\phantom{-}0.200]^{*}$ & $[-0.015,$ & $\phantom{-}0.307]^{\phantom{*}}$ & 4 & 78 & 19 \\
Credit demand (D) & sd & $[-0.088,$ & $\phantom{-}0.052]^{\phantom{*}}$ & $[-0.147,$ & $\phantom{-}0.102]^{\phantom{*}}$ & $[-0.136,$ & $\phantom{-}0.117]^{\phantom{*}}$ & $[-0.201,$ & $\phantom{-}0.223]^{\phantom{*}}$ & 5 & 64 & 31 \\
\bottomrule
\end{tabular}